\documentclass[12pt,preprint]{aastex63}
\usepackage{longtable}
\usepackage{graphicx}
\usepackage{txfonts}
\usepackage{verbatim}
\usepackage{hyperref}
\usepackage{color}
\usepackage{threeparttablex}

\newcommand{\lsun}{log ($L/L_{\odot})\,$}
\newcommand{\msun}{$M/M_{\odot}\,$}
\received{2019 November 14}
\accepted{2020 January 29}
\submitjournal{ApJS}
\shorttitle{Extended theoretical scenario for 
Classical Cepheids}
\shortauthors{G. De Somma et al. }
\graphicspath{{./}{figures/}}

\begin{document}

\title{An extended theoretical scenario for 
Classical Cepheids. I. Modeling Galactic Cepheids in the \textsl{Gaia} photometric system}
\correspondingauthor{Giulia De Somma}
\email{giulia.desomma@inaf.it , gdesomma@na.infn.it}

\author{Giulia De Somma}
\affiliation{Dipartimento di Fisica ”E. Pancini”, Universit`a di Napoli ”Federico II”\\
Compl. Univ. di Monte S. Angelo, Edificio G, Via Cinthia\\
I-80126, Napoli, Italy}
\affiliation{ INAF-Osservatorio astronomico di Capodimonte \\
Via Moiariello 16 \\
80131 Napoli, Italy}
\affiliation{Istituto Nazionale di Fisica Nucleare (INFN) Sez. di Napoli\\
Compl. Univ.di Monte S. Angelo, Edificio G, Via Cinthia \\
I-80126, Napoli, Italy}

\author{Marcella Marconi}
\affiliation{ INAF-Osservatorio astronomico di Capodimonte \\
Via Moiariello 16 \\
80131 Napoli, Italy}

\author{Roberto Molinaro}
\affiliation{ INAF-Osservatorio astronomico di Capodimonte \\
Via Moiariello 16 \\
80131 Napoli, Italy}

\author{Michele Cignoni}
\affiliation{ Dipartimento di Fisica, Università di Pisa \\ 
Largo Bruno Pontecorvo, 3 \\
I-56127 Pisa, Italy}
\affiliation{ INAF-Osservatorio di Astrofisica e Scienza dello Spazio \\
Via Gobetti 93/3 \\
I-40129 Bologna, Italy}
\affiliation{ Istituto Nazionale di Fisica Nucleare (INFN), Sezione di Pisa \\ Largo Pontecorvo 3 \\
I-56127 Pisa, Italy}

\author{Ilaria Musella}
\affiliation{ INAF-Osservatorio astronomico di Capodimonte \\
Via Moiariello 16 \\
80131 Napoli, Italy}

\author{Vincenzo Ripepi}
\affiliation{ INAF-Osservatorio astronomico di Capodimonte \\
Via Moiariello 16 \\
80131 Napoli, Italy}

\begin{abstract}
\noindent
We present a new extended and detailed set of models for Classical Cepheid pulsators at solar chemical
composition ($Z=0.02$, $Y=0.28$) based on a well tested nonlinear hydrodynamical
approach.
In order to model the possible dependence on crucial assumptions such
as the Mass-Luminosity relation of central Helium burning
intermediate-mass stars or the efficiency of superadiabatic convection, the model set
was computed by varying not only the pulsation mode and the stellar mass but also the Mass-Luminosity relation and the mixing length parameter that is used to close the
system of nonlinear hydrodynamical and convective equations.
The dependence of the predicted boundaries of the instability strip
as well as of both light and radial velocity
curves on the assumed Mass-Luminosity and the efficiency of superadiabatic convection is
discussed. Nonlinear Period-Mass-Luminosity-Temperature, Period-Radius and Period-Mass-Radius relations are also computed. 
The theoretical atlas of bolometric light curves for both
the fundamental and first overtone mode  has been
converted in the \textsl{Gaia} filters $G$, $G_{BP}$ and $G_{BR}$ and the
corresponding  mean magnitudes have been derived. Finally the first theoretical Period-Luminosity-Color and Period-Wesenheit relations in the \textsl{Gaia} filters are provided and the resulting theoretical parallaxes are compared with \textsl{Gaia Data Release 2} results for both fundamental and first overtone Galactic Cepheids.

\end{abstract}


\section{Introduction} \label{sec:intro}
Classical Cepheids (CC) are pulsating intermediate-mass central
Helium burning stars
associated to the blue loop phase in the Color-Magnitude diagram.
Their characteristic Period-Luminosity (PL) and
Period-Luminosity-Color (PLC) relations make these objects excellent
distance indicators in the Local Group and beyond. Thanks to the
capability of the Hubble Space Telescope \citep[HST, see e.g.][]{Freedman2001,Riess2011} the Cepheid distance scale
has been extended up to almost 30 Mpc  and still longer distances will
be covered with the next generation observetional facilities, such as the Extremely
Large Telescope from the ground or the James
Webb Space Telescope from the space. 

From the calibration of secondary distance indicators based on
Cepheids in the Milky Way, the LMC, and the maser HST galaxy NGC 4258, \citet{riess2019} derived a
value of the Hubble Constant ($H_{0}= $ 74.03$\pm$ 1.42 $ km s^{-1} Mpc^{-1}$) that is systematically
higher than the value ($H_{0}= $ 67.4$\pm$0.5 $ km s^{-1} Mpc^{-1}$) based on the investigation of the Cosmic
Microwave Background \citep[CMB,][]{planck2018}.
The significant discrepancy between the two estimated values of $H_{0}$ is known as 'The Hubble Constant tension'. In this context it is worth investigating possible residual sources of uncertainties
affecting the Cepheid-based extragalactic distance scale.
We know that Cepheid PL and PLC relations may be affected by
metallicity corrections. Even if this effect has been accounted for in
Riess et al. derivation, several authors provide different
metallicity corrections (e.g. \citet {macri2006,romaniello2008,marconi2005}), in
some cases partially balanced by theoretically predicted Helium
abundance effects (e.g. \citet{carini2014,marconi2005})
and there is no general consensus in the literature.
But even neglecting the metallicity problem, the theory of stellar
evolution and pulsation predicts that other effects can change the
coefficients of the above mentioned relations. These are for example
the efficiency of superadiabatic convection, that contributes to the
quenching of pulsation which affects the topology of the instability strip
as well as the pulsation amplitudes, or the actual Mass-Luminosity
(ML) relation predicted by stellar evolution models for CC which is well known to depend on nonstandard physical
phenomena such as core-overshooting, mass loss and rotation.

In order to quantify these effects on the predicted Cepheid distance
scale we started a theoretical project aiming at building a
complete grid  of nonlinear convective pulsation models
spanning simultaneously a range of possible ML
relations, superadiabatic convection efficiencies and
chemical compositions. The final goal is to provide an extensive and
detailed pulsation model database to complement similar sets of
evolutionary models (see e.g. BaSTI database\footnote{\url{http://basti.oa-teramo.inaf.it/index.html}}, PISA stellar models\footnote{\url{http://astro.df.unipi.it/stellar-models/index.php?m=3}}, 
Padova database of stellar evolutionary tracks and isochrones\footnote{\url{http://pleiadi.pd.astro.it/}}) available to the astrophysical community.

This paper, devoted to the first completed model set at  solar
chemical composition
(Z=0.02 Y=0.28), represents the first step in this direction, whereas
the extension to other chemical compositions will be presented in a
forthcoming work (De Somma et al. in prep).

In this context, the \textsl{Gaia} mission \citep{prusti2016} is producing a 3D-map of 1 billion stars of the Milky Way with unprecedented accuracy.
Focusing on pulsating stars, after the recent Data Release 2 \citep{brown2018,holl2018}, a large sample of Cepheids, observed in three
photometric bands ($G$, $G_{BP}$ and $G_{RP}$) complemented with accurate parallaxes
and proper motions, is available to the scientific community \citep{clementini2019,rip2019}, and the
future releases will provide also radial velocity time series. This important database represents a challenging benchmark for testing the physical and numerical assumptions of current pulsation models. On this basis, in this paper we provide the first set of predicted light curves in the \textsl{Gaia} filters together with the associated mean magnitudes and colors and in turn the inferred 
PLC and Period-Wesenheit (PW) relations in the \textsl{Gaia}
photometric system.

The organization of the paper is the following: in Section 2 we
present the set of pulsation models; in Section 3 we derive the
pulsation relation connecting the period to the intrinsic stellar
parameters, the predicted instability strip
and the theoretical atlas of light and radial velocity curves,
including the effects of the assumed ML and superadiabatic convective efficiency. Moreover, we estimate the Period-Radius (PR) and Period-Mass-Radius
(PMR) relations as a function of the above mentioned assumptions and make a
comparison with independently derived PR relations in the literature;
in Section 4 we provide the first theoretical light curves in the \textsl{Gaia} photometric system from which we obtain mean magnitudes and colors and the first PLC
and the PW relations in the \textsl{Gaia} filters; in Section 5 we derive theoretical parallaxes based on the PW relations in the \textsl{Gaia} filters and make a comparison between theoretical and \textsl{Gaia Data Release 2} (DR2) parallaxes; in Section 6 the conclusions close the paper.

\section{The extended set of pulsation models}
In order to compute the extended set of
Cepheid nonlinear convective pulsation models we adopte the
hydrodynamical code and the physical and numerical assumptions
discussed in \citet{bonotornambe2000,bcm2000,mmb2013, mmfc2010}; but a new automatized procedure
has been developed to compute extended model sets, with unprecedented fine input parameter grids.
These models have solar metallicity $Z = 0.02$ and  Helium content $Y
= 0.28$ and span a wide range of masses (3 $\le$ $\frac{M}{M_{\odot}}$
$\le$ 11) and temperatures ($3600 $K $\le$ $ T_{eff}$ $\le$ $6700 K$). For
each selected stellar mass, three luminosity levels are considered: a
canonical level (named A), based on stellar evolution predictions that neglect
mass loss, rotation and core overshooting \citep{bcm2000} and two noncanonical
luminosity levels obtained by increasing
the canonical luminosity level by 0.2 dex (named B) and 0.4 dex (named C), respectively.
Moreover, in order to investigate the effect of superadiabatic convection efficiency, whose known main effect is to quench the pulsation driving mechanism,
each selected model is computed for three values of the mixing length
parameter $\alpha~$=$~l/H_{P}$\footnote{$\alpha~$=$~l/H_{P}$ where l is the length of the
path covered by the convective elements and $H_{P}$ is the pressure height scale.}
adopted to close the system of nonlinear hydrodynamical and
convective equations \citep{bms1999}, namely $\alpha=1.5$ , $\alpha=1.7$  and $\alpha=1.9$.
The choice of the mixing length parameter range was based on specific computations presented in previous papers (\citet{dicriscienzo2004}; \citet{fmm2007}; \citet{natale2008}; \citet{mmb2013}) which suggested that hotter variables are well reproduced with $\alpha$ = 1.5-1.6, whereas variables closer to the red edge of the instability strip often enquire  $\alpha$ = 1.8-2.0 due to the most important efficiency of convection in the redder part of the color-magnitude diagram.

For each pulsation model the nonlinear equations are integrated until
a stable limit cycle is attained in the F or FO mode.
Table \ref{f_param_model} and Table \ref{fo_param_model} report the intrinsic stellar parameters for the computed F and FO models, respectively. Columns from 1 to 5 report the stellar mass, the luminosity level, the effective temperature, the adopted mixing length parameter, the luminosity level identification defined above. The pulsation period and the average radius inferred from the application of the nonlinear convective code are listed in the last 2 columns.

\begin{ThreePartTable}
\begin{TableNotes}
\footnotesize 
\item[a] Stellar mass (solar units).
\item[b] Logarithmic luminosity (solar units).
\item[c] Effective temperature (K).
\item[d] Mixing length parameter.
\item[e] Mass-Luminosity relation.
\item[f] Period (days).
\item[g] Logarithmic mean radius (solar units).
\end{TableNotes}
\begin{longtable}{ccccccc}
\caption{\label{f_param_model} The intrinsic stellar parameters for computed F mode models. Full tables are available in the Appendix.}\\
\hline\hline
&&&Z=0.02 & Y= 0.28\\
M\tnote{a}&logL\tnote{b}&$T_{eff}$\tnote{c}&$\alpha$\tnote{d}&ML\tnote{e}&P\tnote{f}&$log\overline{R}$\tnote{g}\\
\hline
(1)&(2)&(3)&(4)&(5)&(6)&(7)\\
\hline
\endfirsthead
\caption{continued.}\\
\hline\hline
M\tnote{a}&logL\tnote{b}&$T_{eff}$\tnote{c}&$\alpha$\tnote{d}&ML\tnote{e}&P\tnote{f}&$log\overline{R}$\tnote{g}\\
\hline
(1)&(2)&(3)&(4)&(5)&(6)&(7)\\
\hline
\endhead
\hline
\endfoot
3.0&2.32&5900&1.5&A&1.07716&1.142\\
3.0&2.32&6000&1.5&A&1.03611&1.129\\
...\\
4.0&2.74&5500&1.5&A&2.56311&1.412\\
4.0&2.74&5600&1.5&A&2.42218&1.399\\
...\\
5.0&3.07&5300&1.5&A&4.73277&1.608\\
5.0&3.07&5400&1.5&A&4.44069&1.592\\
...\\
6.0&3.33&5000&1.5&A&8.6011&1.789\\
6.0&3.33&5100&1.5&A&8.0714&1.772\\
...\\
7.0&3.56&4800&1.5&A&14.00799&1.942\\
7.0&3.56&4900&1.5&A&13.0515&1.928\\
...\\
8.0&3.75&4600&1.5&A&21.7684&2.070\\
8.0&3.75&4700&1.5&A&20.2235&2.056\\
...\\
9.0&3.92&4400&1.5&A&33.08715&2.190\\
9.0&3.92&4500&1.5&A&30.575&2.174\\
...\\
10.0&4.08&4200&1.5&A&48.711&2.297\\
10.0&4.08&4300&1.5&A&45.74965&2.285\\
...\\
11.0&4.21&4100&1.5&A&66.40289&2.386\\
11.0&4.21&4200&1.5&A&61.14294&2.371\\
...\\
\hline
\insertTableNotes   
\end{longtable}
\end{ThreePartTable}

\begin{ThreePartTable}
\begin{TableNotes}
\footnotesize 
\item[a] Stellar mass (solar units).
\item[b] Logarithmic luminosity (solar units).
\item[c] Effective temperature (K).
\item[d] Mixing length parameter.
\item[e] Mass-Luminosity relation.
\item[f] Metal abundance.
\item[i] Logarithmic mean radius (solar units).
\end{TableNotes}
\begin{longtable}{ccccccc}
\caption{\label{fo_param_model} The intrinsic stellar parameters for computed FO mode models. Full tables are available in the Appendix.}\\
\hline\hline
&&&Z=0.02 & Y= 0.28\\
M\tnote{a}&logL\tnote{b}&$T_{eff}$\tnote{c}&$\alpha$\tnote{d}&ML\tnote{e}&P\tnote{f}&$log\overline{R}$\tnote{g}\\
\hline
(1)&(2)&(3)&(4)&(5)&(6)&(7)\\
\hline
\endfirsthead
\caption{continued.}\\
\hline\hline
M\tnote{a}&logL\tnote{b}&$T_{eff}$\tnote{c}&$\alpha$\tnote{d}&ML\tnote{e}&P\tnote{f}&$log\overline{R}$\tnote{g}\\
\hline
(1)&(2)&(3)&(4)&(5)&(6)&(7)\\
\hline
\endhead
\hline
\endfoot
3.0&2.32&6200&1.5&A&0.6715&1.103\\
3.0&2.32&6300&1.5&A&0.6403&1.090\\
...\\
4.0&2.74&5900&1.5&A&1.4240&1.354\\
4.0&2.74&6000&1.5&A&1.3551&1.341\\
...\\
5.0&3.07&5800&1.5&A&2.3904&1.530\\
5.0&3.07&5900&1.5&A&2.2912&1.517\\
...\\
6.0&3.33&5800&1.5&A&3.5712&1.664\\
\hline
\insertTableNotes
\end{longtable}
\end{ThreePartTable}
 
\section{Results from the extended model set}
In this section we present the theoretical predictions obtained from the extended grid of models, concerning the period dependence on the intrinsic stellar parameters, the instability strip, the bolometric light and radial velocity curves and PR and PMR relations, as a function of both the ML relation and $\alpha$ value.

\subsection{The Period-Luminosity-Mass-Temperature relations}
We carried out a linear regression analysis of the values reported in Tables \ref{f_param_model} and \ref{fo_param_model} to obtain the pulsation relations connecting the Period to the
Luminosity, the Mass and the Effective Temperature (PLMT) for both the
F and FO models as a function of the assumed mixing length
parameter.
The coefficients are reported in Table \ref{pmlt_f_fo} for the F and FO pulsators, respectively.
These relations, that update previous relations published
in the literature for solar metallicity models \citep{bcm2000} are consistent with the latter, within the errors, and confirm the use of pulsation models to establish sound relations between pulsational and evolutionary parameters. Moreover they provide a very important tool to build iso-periodic model sequences to apply the light and radial velocity curves model fitting \citep{mmb2013,m2017}.
The coefficients reported in previous Table \ref{pmlt_f_fo} show that a variation of the
mixing length parameter does not significantly affect the PLMT
relations. This result is expected because the PMLT relation directly derives from the combination of the Period-Mean Density relation and the Stefan-Boltzmann law and holds for each individual pulsator indipendently of the position in the Hertzprung-Russel (HR) diagram. However, for $\alpha=1.9$ the number of pulsating models is significantly decreased and limited to the lower masses, thus affecting the shape of the relations.

\begin{ThreePartTable}
\begin{longtable}{cccccccccc}
\caption{\label{pmlt_f_fo} Coefficients of the F and FO pulsators PMLT relations $\log P$ = a +b$\log T_{eff}$ + c $\log$ (\msun) + d \lsun as a function of the assumed $\alpha$ parameter.}\\
\hline\hline
$\alpha$&a&b&c&d& $\sigma_{a}$& $\sigma_{b}$& $\sigma_{c}$& $\sigma_{d}$&$R^2$\\
\hline
F\\
\hline
\endfirsthead
\hline\hline
$\alpha$&a&b&c&d& $\sigma_{a}$& $\sigma_{b}$& $\sigma_{c}$& $\sigma_{d}$&$R^2$\\
\hline
\endhead
\hline
\endfoot
1.5&10.268&-3.192&-0.758&0.919&0.001&0.025&0.015&0.005&0.9995\\
1.7&10.538&-3.258&-0.749&0.911&0.002&0.050&0.019&0.007&0.9996\\
1.9&11.488&-3.469&-0.695&0.847&0.003&0.089&0.012&0.006&0.9999\\
\hline
FO\\
1.5&10.595&-3.253&-0.621&0.804&0.002&0.067&0.014&0.005&0.9996\\
1.7&10.359&-3.186&-0.576&0.788&0.002&0.056&0.009&0.003&0.9999\\
\hline
\end{longtable}
\end{ThreePartTable}

\subsection{The new predicted instability strip}
In this subsection we present the variation of the topology of the
instability strip obtained for both F and FO models as we change the
ML relation (from case A to case B and C) and the efficiency of superadiabatic convection from $\alpha=1.5$ to 1.7 and 1.9.
The stability of both pulsation modes is investigated in order to
predict the hottest and the coolest model for each combination of
$M$, $L$ and $\alpha$ . The blue and red boundaries of the F and FO strips
are then evaluated by increasing/decreasing by 50 K the effective temperature of the bluest/reddest model.

\begin{figure}[ht!]
\centering
\includegraphics[width=0.5\textwidth]{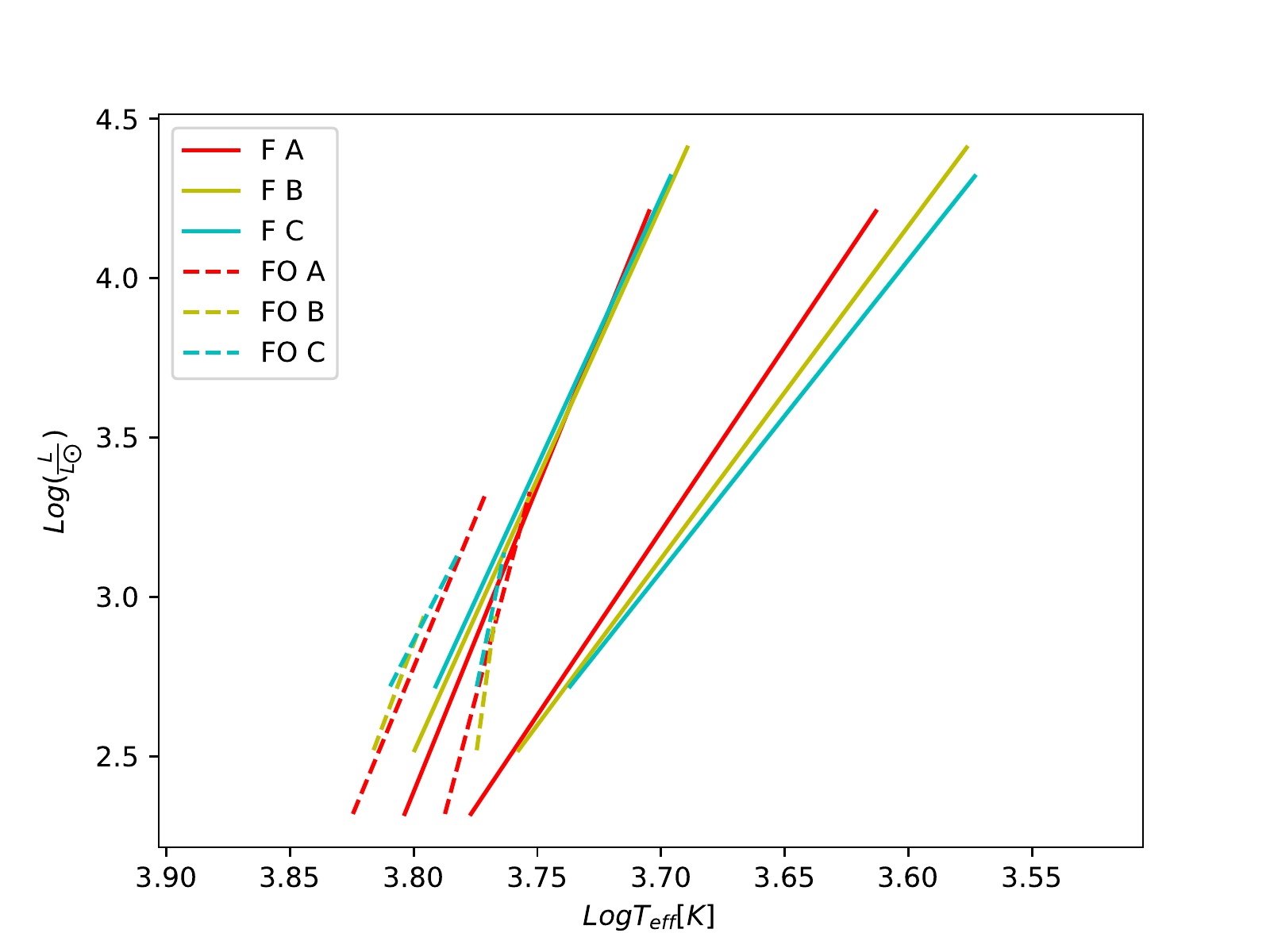}
\caption{F and FO instability strips at fixed mixing length parameter $\alpha=1.5$ for the assumed A, B, C ML relations.}
\label{Fig:strip_ml}
\end{figure}

\begin{figure}[ht!]
\centering
\includegraphics[width=0.5\textwidth]{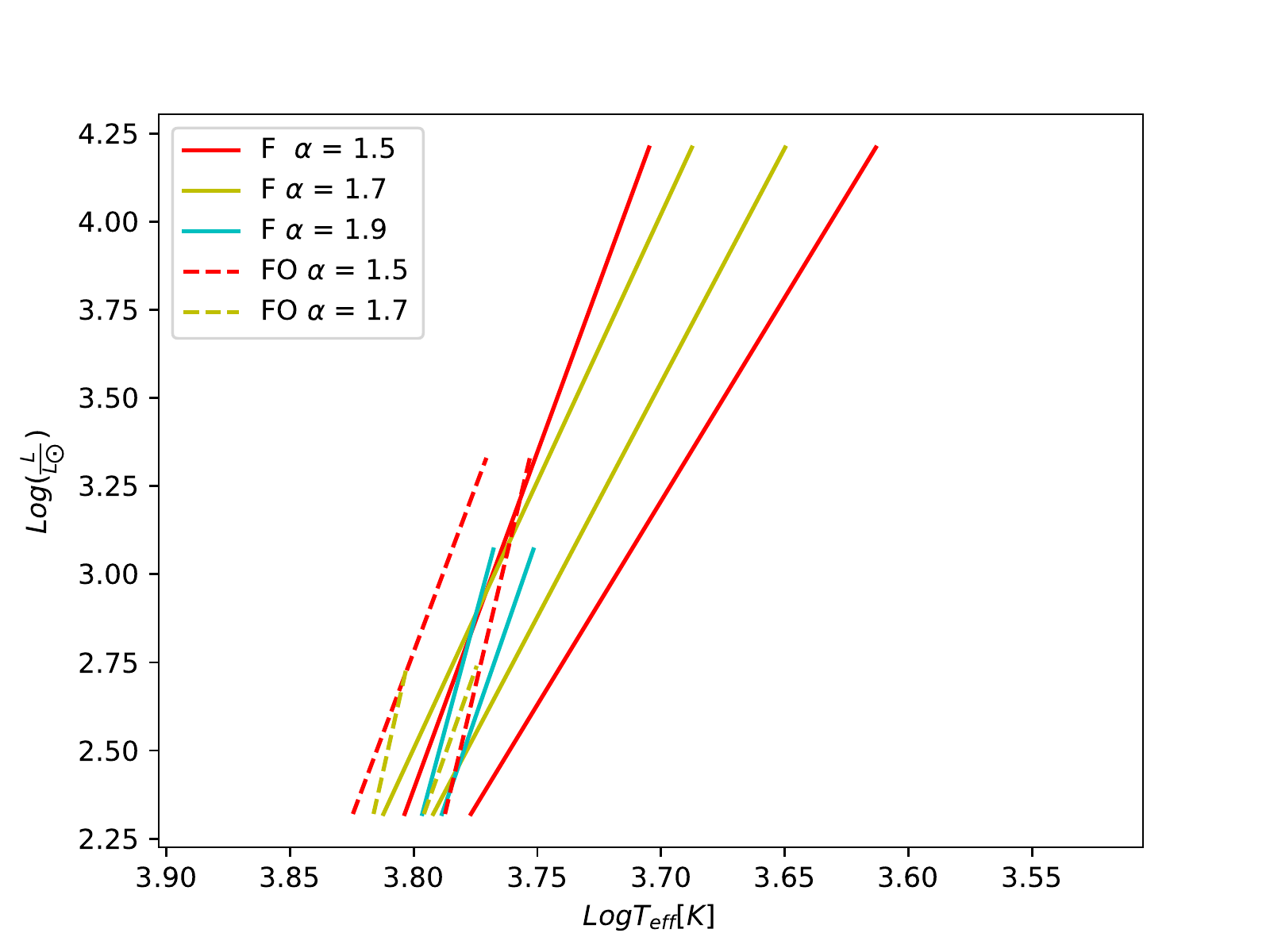}
\caption{Canonical F and FO instability strips for the various assumption about superadiabatic convective efficiency.}
\label{Fig:strip_alfa}
\end{figure}

The resulting boundaries are reported in Table 4. Columns from 1 to
8 provide the mass, the luminosity level, the adopted mixing length parameter, the ML label, the first overtone blue edge (FOBE), the fundamental blue edge (FBE), the first overtone red edge (FORE) and the fundamental red edge (FRE).
We notice that the FO pulsation is found only for masses lower than 6$M_\odot$ in agreement with previous results in the literature \citep{bcm2000}.
Linear regression through the values reported in Table \ref{boundaries}
for the FO and F boundaries respectively allows us to derive the
relations reported in Table \ref{pl_f} and Table \ref{pl_fo} again at varying the ML relation and $\alpha$ . We note that whilst the majority of the $R^2$  values of these regressions are above 0.9, for the case of our brightest ML (case C) the FRE relations seem to be less accurate. This occurrence reflects the trend, already discussed in some previous papers \citep{bcm2000}, of the FRE getting hotter when the brightest luminosity levels are achieved as a consequence of the decreased density in the driving regions.
These relations are plotted in Figures (\ref{Fig:strip_ml}) and (\ref{Fig:strip_alfa}) at varying the ML relation and the superadiabatic convection efficiency, respectively. Inspection of these plots suggests that, in agreement with previous theoretical indications \citep{bcm2000,fmm2007}, a variation in the ML relation (Figure \ref{Fig:strip_ml}) does not significantly affect the topology of the instability strip, whilst increasing the efficiency of superadiabatic convection implies a quenching effect on pulsation and in turn a narrowing of the instability strip, in agreement with previous investigations (e.g.\citet{fmm2007}). In particular, an increase in the  $\alpha$ parameter (Figure \ref{Fig:strip_alfa}) makes the FOBE redder by about $100 K$ and the FRE bluer  by about $300 K$ confirming that the quenching effect due to superadiabatic convection is particularly efficient in the red part of the instability strip.

\begin{ThreePartTable}
\begin{TableNotes}
\footnotesize 
\item[a] Stellar mass (solar units).
\item[b] Logarithmic luminosity (solar units).
\item[c] Mixing length parameter.
\item[d] Mass-Luminosity relation.
\item[e] First overtone blue edge.
\item[f] Fundamental blue edge.
\item[g] First overtone red edge.
\item[h] Fundamental red edge.
\end{TableNotes}
\begin{longtable}{cccccccc}
\caption{\label{boundaries} Predicted Effective Temperatures of the Instability Strip Boundaries. The assumed error on the predicted  boundaries of the instability strip is $\pm$ 50 K, as based on our assumed effective temperature step in  the building of the pulsation model grid.}\\
\hline\hline
M\tnote{a}&logL\tnote{b}&$\alpha$\tnote{c}&ML\tnote{d}&FOBE\tnote{e}&FBE\tnote{f}&FORE\tnote{g}&FRE\tnote{h}\\
\hline
(1)&(2)&(3)&(4)&(5)&(6)&(7)&(8)\\
\hline
\endfirsthead
\caption{continued.}\\
\hline\hline
M\tnote{a}&logL\tnote{b}&$\alpha$\tnote{c}&ML\tnote{d}&FOBE\tnote{e}&FBE\tnote{f}&FORE\tnote{g}&FRE\tnote{h}\\
\hline
(1)&(2)&(3)&(4)&(5)&(6)&(7)&(8)\\
\hline
\endhead
\hline
\endfoot
3.0&2.32&1.5&A&6550&6150&6150&5850\\
3.0&2.32&1.7&A&6550&6250&6250&6050\\
3.0&2.32&1.9&A&&6250&&6150\\
3.0&2.52&1.5&B&6550&6050&5950&5550\\
3.0&2.52&1.7&B&6550&6150&6150&5750\\
3.0&2.52&1.9&B&&6150&&5950\\
3.0&2.72&1.5&C&6450&6050&5950&5350\\
3.0&2.72&1.7&C&6250&6150&6150&5550\\
3.0&2.72&1.9&C&&6150&&5750\\
4.0&2.74&1.5&A&6450&5950&5850&5450\\
4.0&2.74&1.7&A&6350&6050&5850&5750\\
4.0&2.74&1.9&A&&6050&&5850\\
4.0&2.94&1.5&B&6250&5950&5850&5250\\
4.0&2.94&1.7&B&&5950&&5450\\
4.0&2.94&1.9&B&&5950&&5650\\
4.0&3.14&1.5&C&6050&5850&5950&4950\\
4.0&3.14&1.7&C&&5850&&5150\\
4.0&3.14&1.9&C&&5750&&5450\\
5.0&3.07&1.5&A&6150&5850&5750&5250\\
5.0&3.07&1.7&A&&5950&&5450\\
5.0&3.07&1.9&A&&5850&&5650\\
5.0&3.27&1.5&B&&5850&&4950\\
5.0&3.27&1.7&B&&5750&&5150\\
5.0&3.47&1.5&C&&5750&&4550\\
5.0&3.47&1.7&C&&5550&&4850\\
5.0&3.47&1.9&C&&5350&&5250\\
6.0&3.33&1.5&A&5850&5850&5750&4950\\
6.0&3.33&1.7&A&&5650&&5250\\
6.0&3.53&1.5&B&&5650&&4650\\
6.0&3.53&1.7&B&&5450&&4950\\
6.0&3.73&1.5&C&&5350&&4250\\
6.0&3.73&1.7&C&&5250&&4550\\
7.0&3.56&1.5&A&&5550&&4750\\
7.0&3.56&1.7&A&&5450&&5150\\
7.0&3.76&1.5&B&&5350&&4350\\
7.0&3.76&1.7&B&&5250&&4750\\
7.0&3.96&1.5&C&&5150&&3950\\
7.0&3.96&1.7&C&&5050&&4350\\
8.0&3.75&1.5&A&&5450&&4550\\
8.0&3.75&1.7&A&&5250&&4850\\
8.0&3.95&1.5&B&&5250&&4150\\
8.0&3.95&1.7&B&&5050&&4450\\
8.0&4.15&1.5&C&&5150&&3750\\
8.0&4.15&1.7&C&&4950&&4050\\
9.0&3.92&1.5&A&&5250&&4350\\
9.0&3.92&1.7&A&&5150&&4750\\
9.0&4.12&1.5&B&&5050&&3950\\
9.0&4.12&1.7&B&&4850&&4250\\
9.0&4.32&1.5&C&&4950&&4150\\
9.0&4.32&1.7&C&&4750&&4250\\
10.0&4.08&1.5&A&&5150&&4150\\
10.0&4.08&1.7&A&&4950&&4550\\
10.0&4.28&1.5&B&&4950&&3750\\
10.0&4.28&1.7&B&&4750&&4050\\
10.0&4.48&1.5&C&&4850&&4450\\
10.0&4.48&1.7&C&&4750&&4450\\
11.0&4.21&1.5&A&&4950&&4050\\
11.0&4.21&1.7&A&&4750&&4350\\
11.0&4.41&1.5&B&&4850&&3950\\
11.0&4.41&1.7&B&&4750&&3850\\
11.0&4.61&1.5&C&&4850&&4550\\
11.0&4.61&1.7&C&&4650&&4550\\
\hline
\insertTableNotes 
\end{longtable}
\end{ThreePartTable}
\clearpage

\begin{ThreePartTable}
\begin{longtable}{ccccccc}
\caption{\label{pl_f} Coefficients of the F pulsator relation $\log T_{eff}$ = a +b \lsun for the various assumptions about the ML relation and the mixing length value.}\\
\hline\hline
$\alpha$&ML&a&b&$\sigma_{a}$&$\sigma_{b}$&$R^2$\\
\hline
FBE\\
\hline
\endfirsthead
\caption{continued.}\\
\hline\hline
$\alpha$&ML&a&b&$\sigma_{a}$&$\sigma_{b}$&$R^2$\\
\hline
\endhead
\hline
\endfoot
1.5&A&3.91&-0.05&0.02&0.005&0.917\\
1.5&B&3.93&-0.05&0.02&0.005&0.944\\
1.5&C&3.94&-0.05&0.01&0.004&0.967\\
1.7&A&3.95&-0.06&0.02&0.005&0.954\\
1.7&B&3.96&-0.06&0.01&0.003&0.976\\
1.7&C&3.97&-0.07&0.009&0.002&0.991\\
1.9&A&3.88&-0.04&0.008&0.003&0.994\\
\hline
FRE\\
\hline
1.5&A&3.97&-0.08&0.01&0.004&0.985\\
1.5&B&3.99&-0.09&0.02&0.006&0.966\\
1.5&C&3.83&-0.05&0.08&0.02&0.434\\
1.7&A&3.96&-0.07&0.015&0.004&0.975\\
1.7&B&4.00&-0.09&0.02&0.006&0.965\\
1.7&C&3.88&-0.06&0.05&0.01&0.707\\
1.9&A&3.90&-0.05&0.004&0.002&0.998\\
\hline
\end{longtable}
\end{ThreePartTable}

\begin{ThreePartTable}
\begin{longtable}{ccccccc}
\caption{\label{pl_fo} Coefficients of the FO pulsator relation $\log T_{eff}$ = a +b\lsun for the various assumptions about the ML relation and the mixing length value.}\\
\hline\hline
$\alpha$&ML&a&b&$\sigma_{a}$&$\sigma_{b}$&$R^2$\\
\hline
FOBE\\
\hline
\endfirsthead
\caption{continued.}\\
\hline\hline
$\alpha$&ML&a&b&$\sigma_{a}$&$\sigma_{b}$&$R^2$\\
\hline
\endhead
\hline
\endfoot
1.5&A&3.94&-0.05&0.03&0.01&0.9045\\
\hline
FORE\\
\hline
1.5&A&3.85&-0.03&0.02&0.008&0.8690\\
\hline
\end{longtable}
\end{ThreePartTable}

\subsection{The light  and radial velocity curves}
\begin{figure}[ht!]
\centering
\textbf{\msun=3.0\,\,\,\,\lsun=2.32\,\,\,\,$\alpha = 1.5$ }\par\medskip
\includegraphics[trim=20 20 0 55,clip,width=0.8\textwidth]{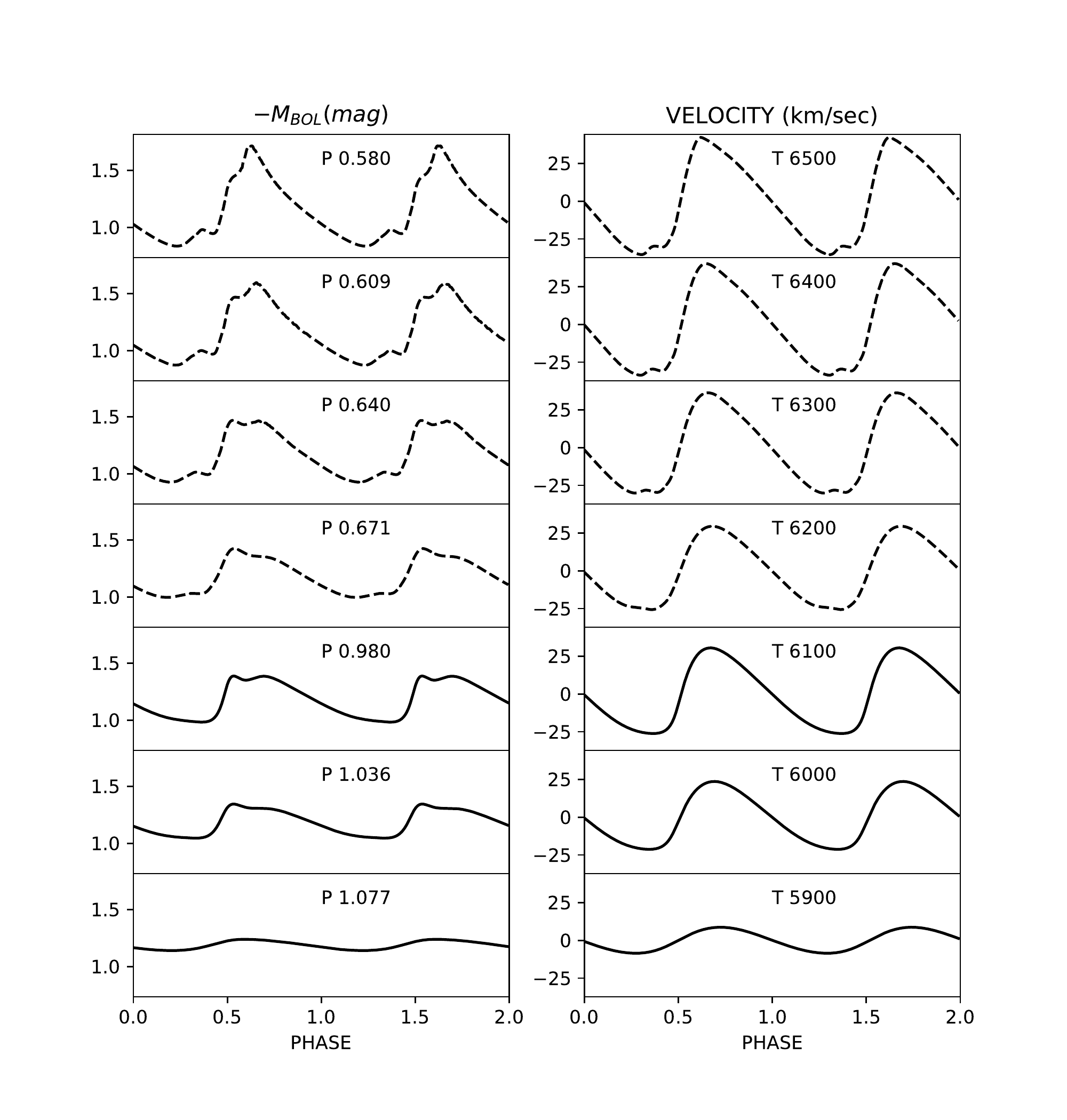}
\caption{Bolometric light curves (left panel) and radial velocity curves (right panel) for a sequence of non linear F (solid line) and FO models (dashed lines) derived at fixed mass, luminosity, $\alpha$ parameter (see labeled values on the top of the plot) adopting the canonical ML relation.}
\label{Fig:lc_vr_2.32_3.0_1.5_ca}
\end{figure}

In this subsection, we present the new theoretical atlas of light and radial velocity curves for both F and FO modes, resulting from the nonlinear computation of full amplitude models.
The predicted bolometric light curves (left panels) and  radial
velocity variations (right panels) are shown in Figure \ref{Fig:lc_vr_2.32_3.0_1.5_ca} for the canonical model sequences. The
curves are plotted over two consecutive pulsation cycles, as a function of the pulsation phase. In each plot, dashed lines refer to FO models, whereas solid lines represent F models. In the left panel, on each light curve the period in days of the model is labeled while in the right panel, on each radial velocity curve the static effective temperature in kelvin is labeled.
The decrease of the amplitudes of both light and radial velocity curves is an expected result related to the quenching effect of convection on pulsation. As $\alpha$ increases the efficiency of superadiabtic convection increases, making the driving mechanisms of pulsation less and less efficient and the amplitude of the oscillation smaller and smaller.
Moreover, positive values of radial velocity along the curves indicate an expansion phase for the stellar envelope, while negative values of the velocity correspond to a contraction phase. The complete atlas of the bolometric light curves for the various assumptions about the ML relation and the superadiabatic convection efficiency are available in the Appendix.

Focusing on canonical models (luminosity level A) with $\alpha=1.5$, we notice that for masses lower or equal than 5 $M_{\odot}$  the curve amplitudes
steadily decrease as the effective temperature decreases, moving from
the FBE  to the FRE.
Above $\sim$ $5M_{\odot}$ this trend is less evident because the FO
pulsation is no  more efficient. In particular in the period range from
$\sim 7$ to $\sim 12$ days a secondary maximum (bump) is present in the
light and radial velocity curves. For this reason, Cepheids in this
period range are called ''bump Cepheids''. The evolution of the bump pulsation phase from the descending to the ascending branch of the curve is the so
called Hertzprung progression (HP) \citep{bms2000}. At the center of the
HP the principal and secondary maximum are very
close in magnitude and the pulsation amplitude often reaches a
minimum. A detailed investigation of the dependence of the period
corresponding to the HP center on metallicity is postponed to a future
paper.

\clearpage

\begin{figure}[ht!]
\centering
\textbf{\msun=4.0\,\,\,\,\lsun=2.74\,\,\,\,$\alpha = 1.5$ }\par\medskip
\includegraphics[trim=20 20 0 95,clip,width=0.8\textwidth]{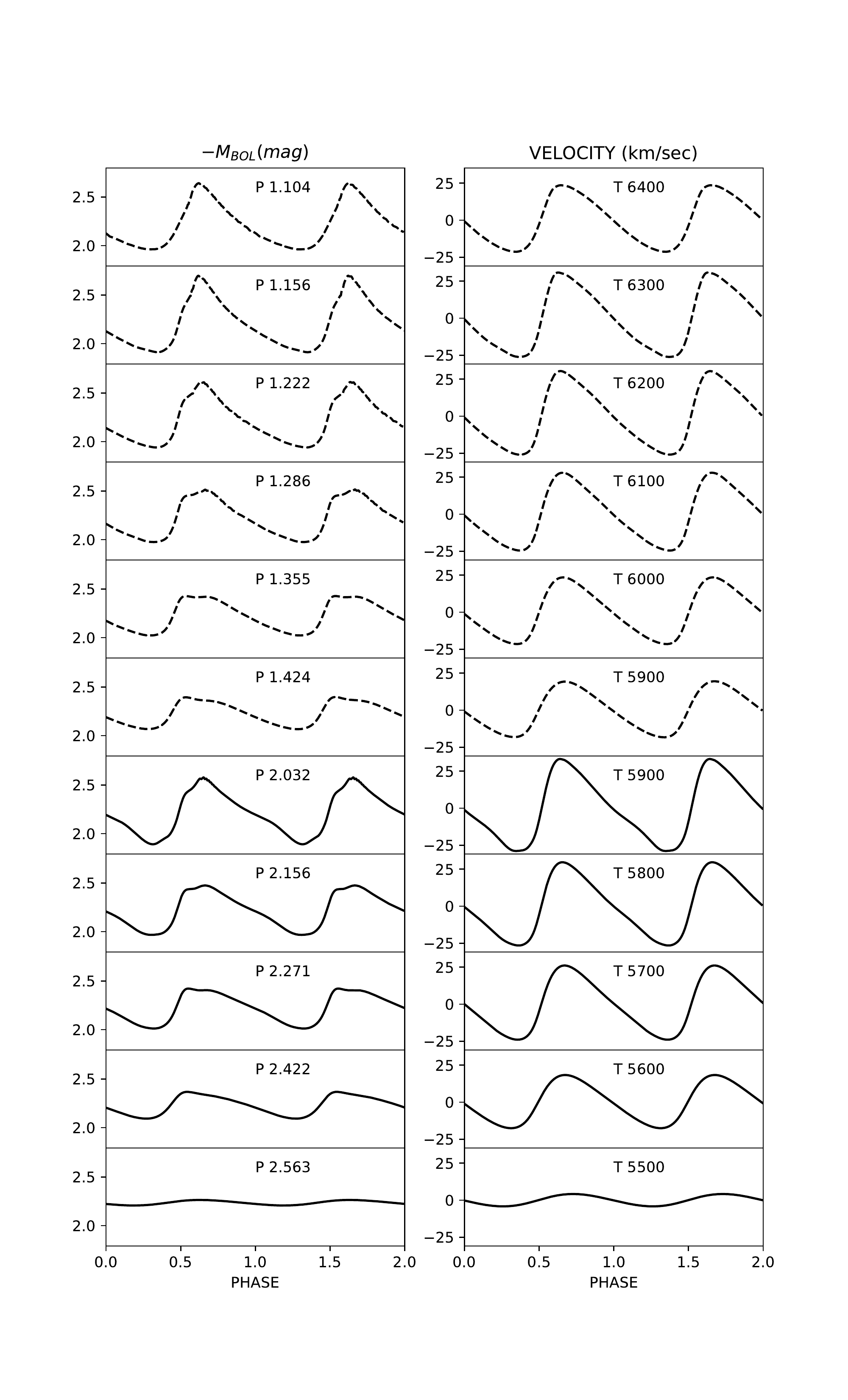}
\par FIG.3-Continued.
\label{Fig:lc_vr_4.0_1.5_ca}
\end{figure}
\clearpage

\begin{figure}[ht!]
\centering
\textbf{\msun=5.0\,\,\,\,\lsun=3.07\,\,\,\,$\alpha = 1.5$ }\par\medskip
\includegraphics[trim=20 20 0 85,clip,width=0.8\textwidth]{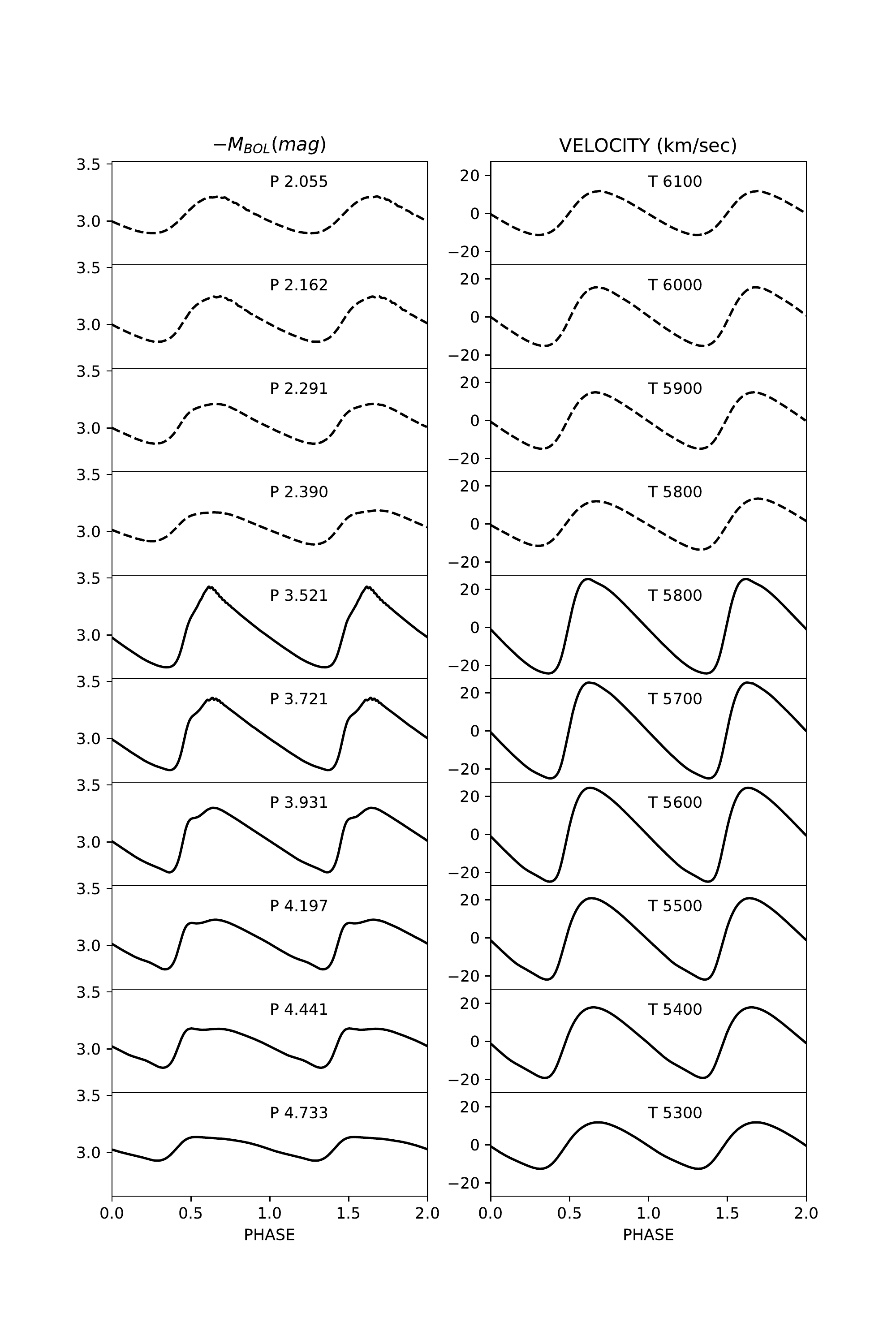}
\par FIG.3-Continued.
\label{Fig:lc_vr_3.07_5.0_1.5_ca}
\end{figure}
\clearpage

\begin{figure}[ht!]
\centering
\textbf{\msun=6.0\,\,\,\,\lsun=3.33\,\,\,\,$\alpha = 1.5$ }\par\medskip
\includegraphics[trim=20 20 0 83,clip,width=0.8\textwidth]{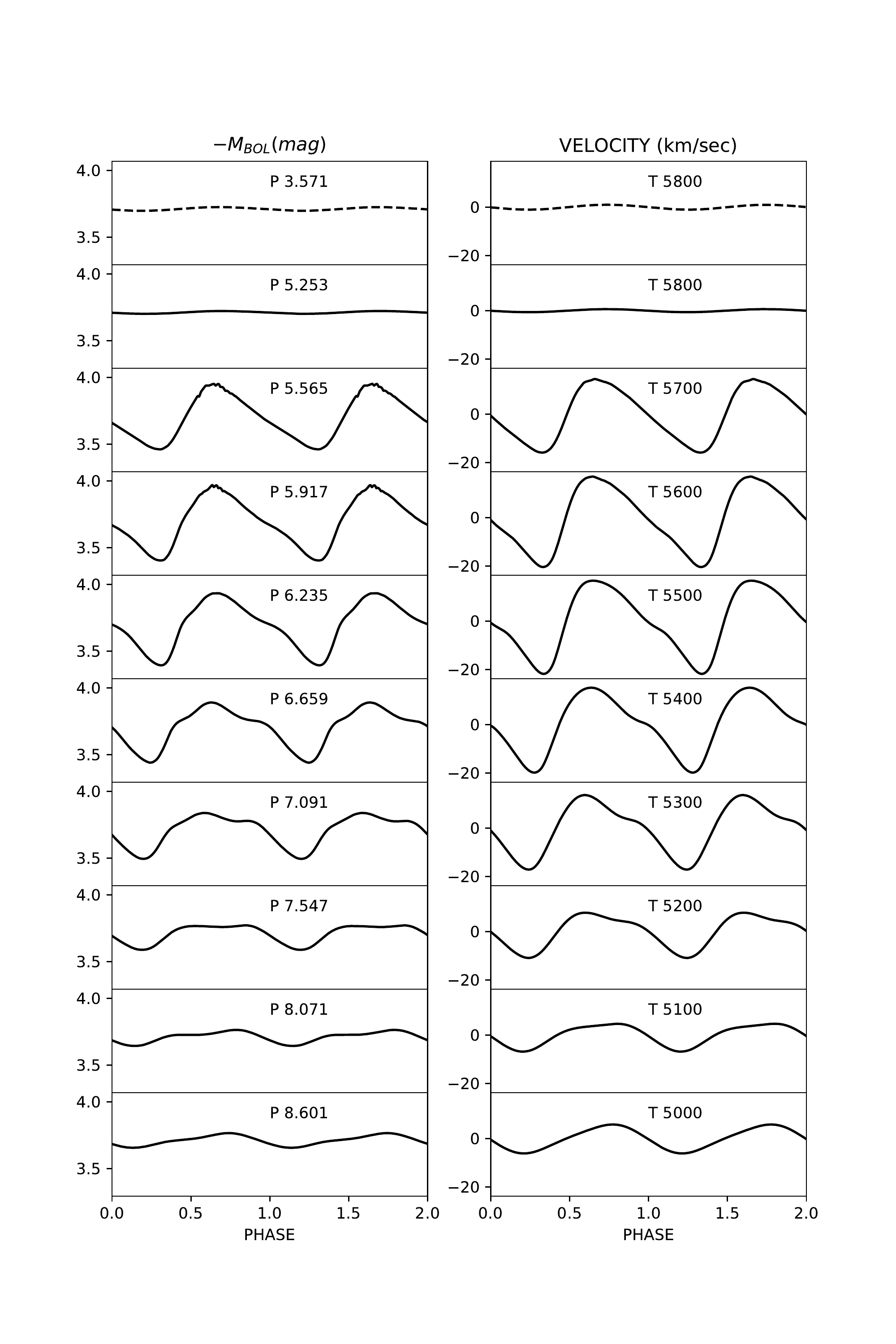}
\par FIG.3-Continued.
\label{Fig:lc_vr_3.33_6.0_1.5_ca}
\end{figure}
\clearpage

\begin{figure}[ht!]
\centering
\textbf{\msun=7.0\,\,\,\,\lsun=3.56\,\,\,\,$\alpha = 1.5$ }\par\medskip
\includegraphics[trim=20 20 0 63,clip,width=0.8\textwidth]{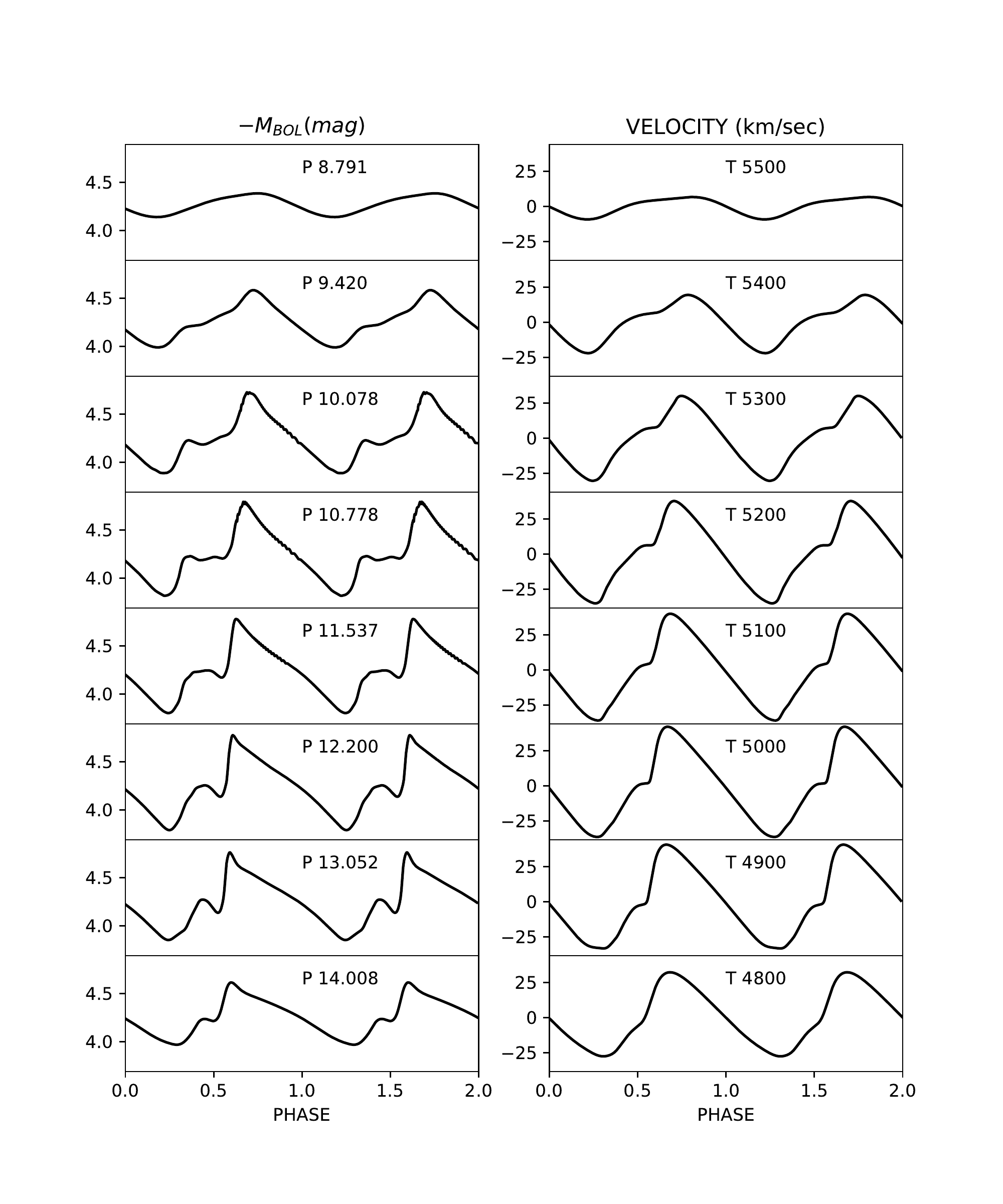}
\par FIG.3-Continued.
\label{Fig:lc_vr_7.0_1.5_ca}
\end{figure}
\clearpage

\begin{figure}[ht!]
\centering
\textbf{\msun=8.0\,\,\,\,\lsun=3.75\,\,\,\,$\alpha = 1.5$ }\par\medskip
\includegraphics[trim=20 20 0 74,clip,width=0.8\textwidth,height=0.95\textheight]{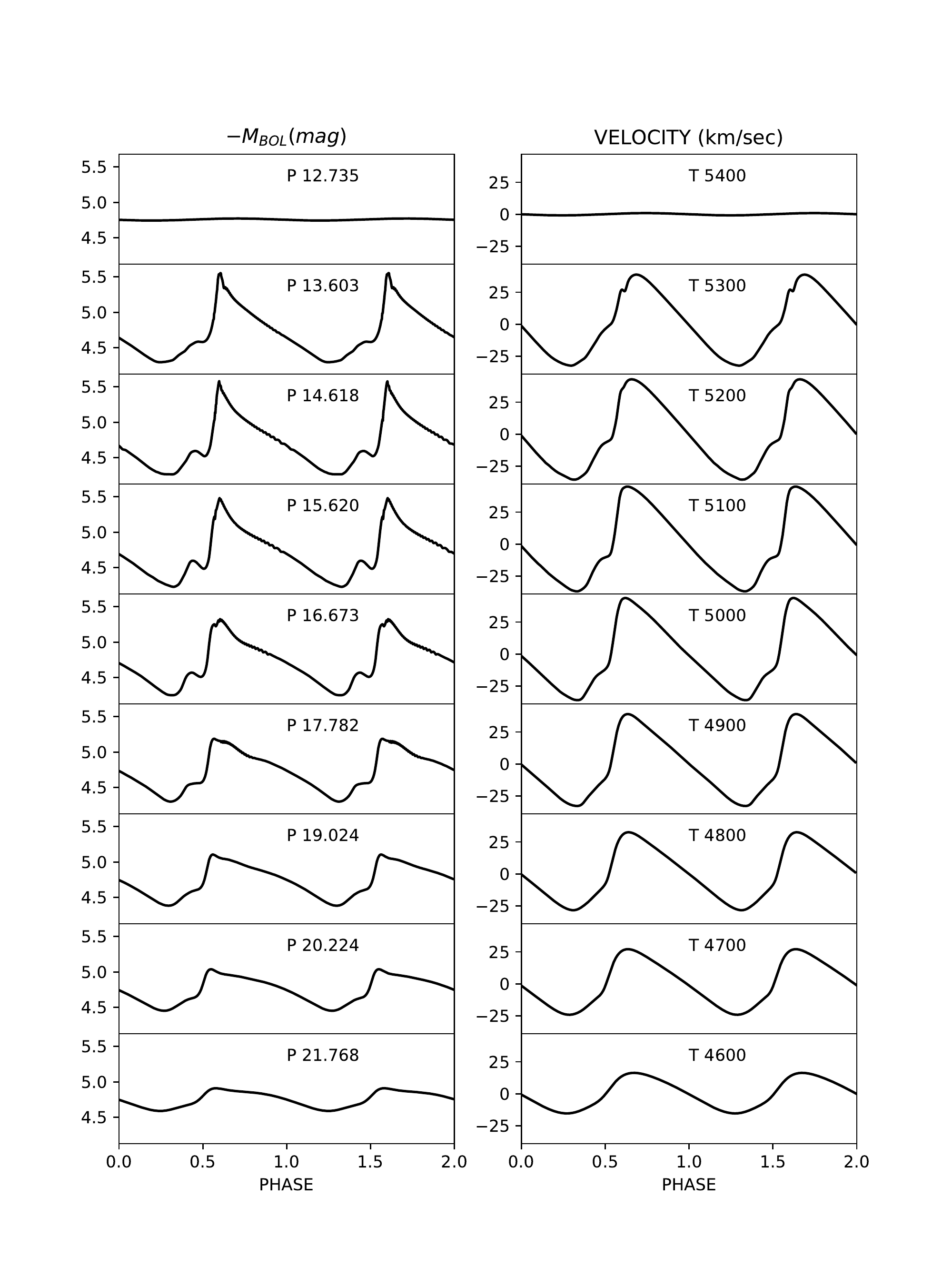}
\par FIG.3-Continued.
\label{Fig:lc_vr_3.75_8.0_1.5_ca}
\end{figure}
\clearpage

\begin{figure}[ht!]
\centering
\textbf{\msun=9.0\,\,\,\,\lsun=3.92\,\,\,\,$\alpha = 1.5$ }\par\medskip
\includegraphics[trim=20 20 0 75,clip,width=0.8\textwidth,height=0.95\textheight]{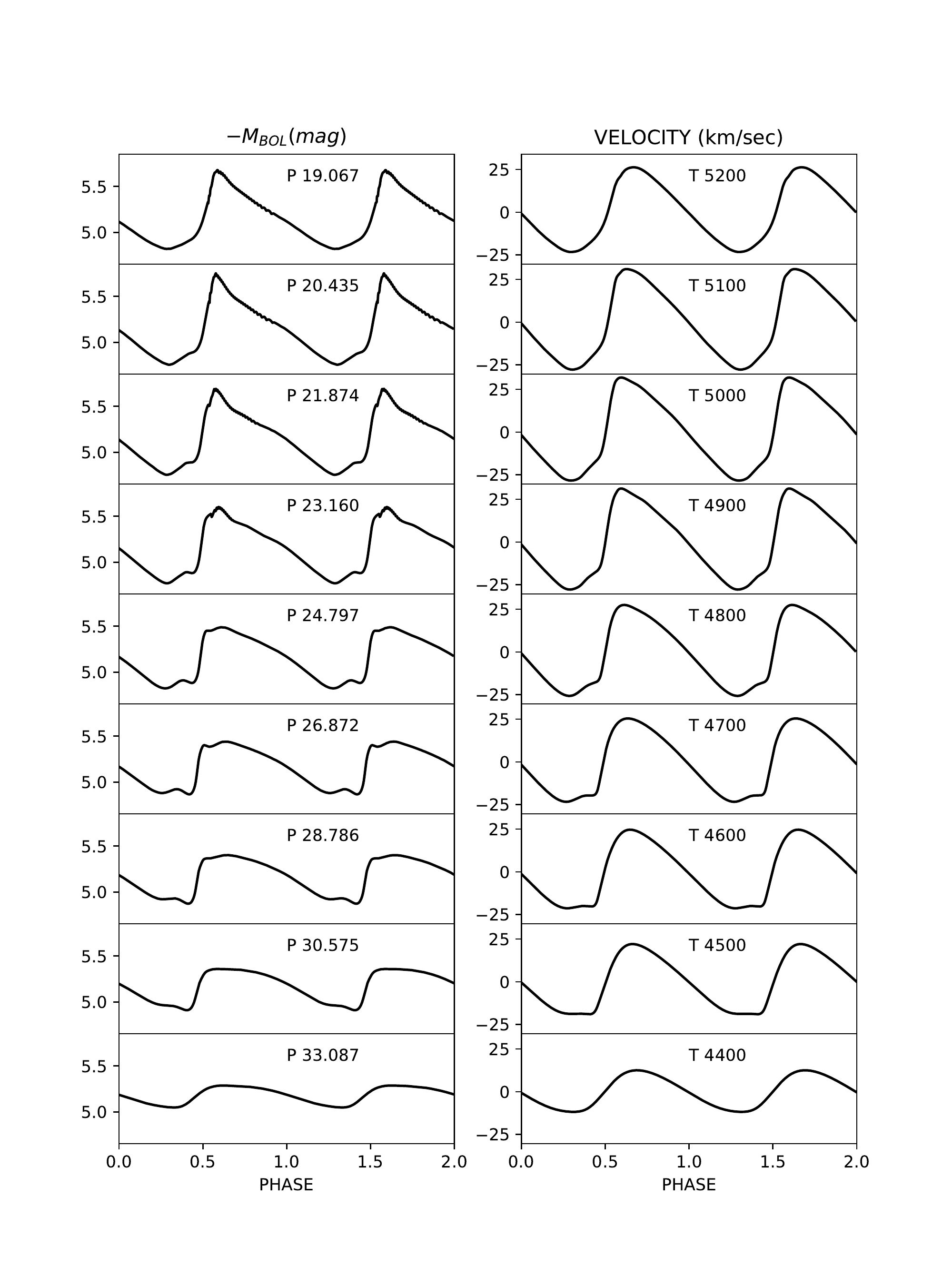}
\par FIG.3-Continued.
\label{Fig:lc_vr_3.92_9.0_1.5_ca}
\end{figure}
\clearpage

\begin{figure}[ht!]
\centering
\textbf{\msun=10.0\,\,\,\,\lsun=4.08\,\,\,\,$\alpha = 1.5$ }\par\medskip
\includegraphics[trim=20 20 0 85,clip,width=0.8\textwidth]{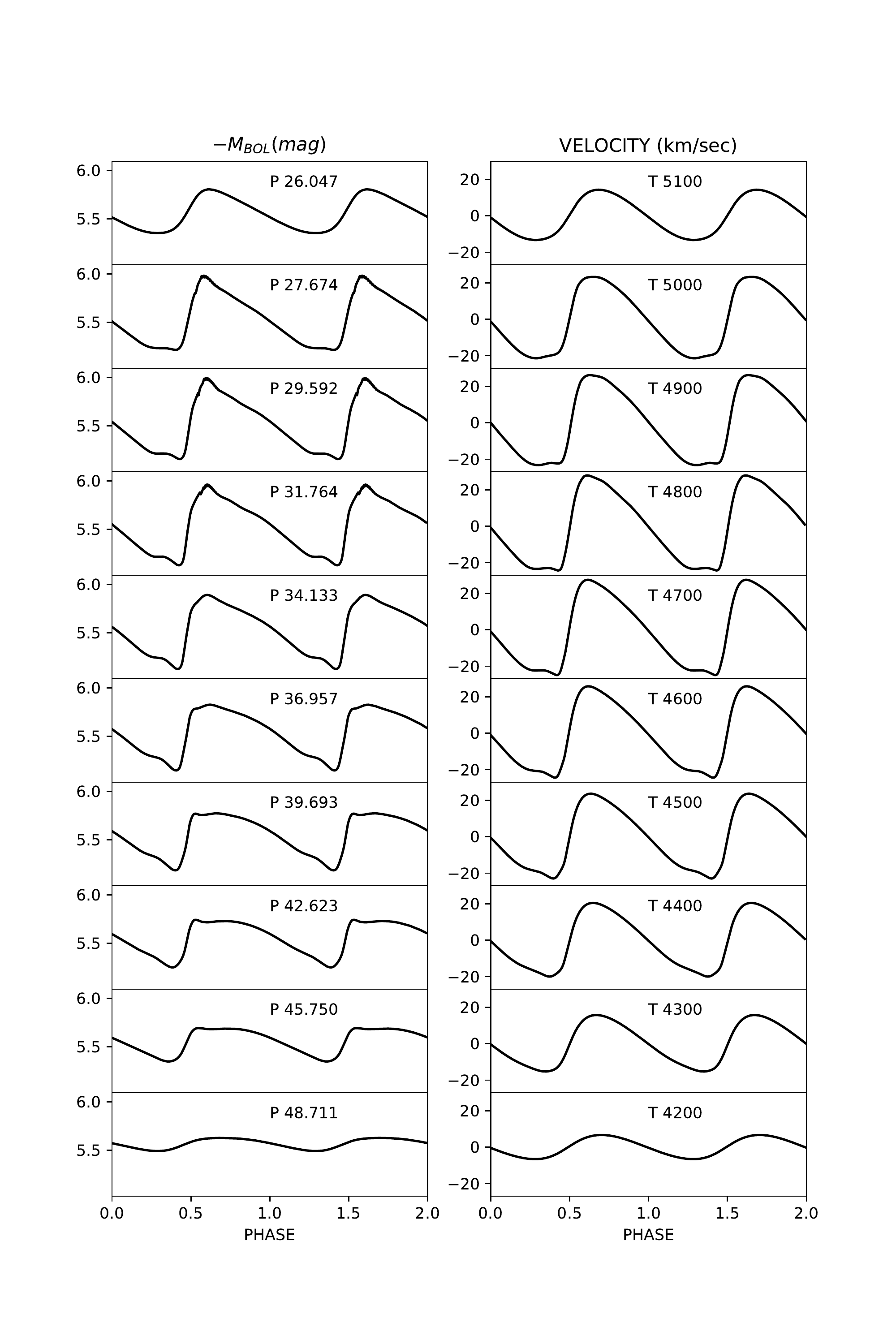}
\par FIG.3-Continued.
\label{Fig:lc_vr_4.08_10.0_1.5_ca}
\end{figure}
\clearpage

\begin{figure}[ht!]
\centering
\textbf{\msun=11.0\,\,\,\,\lsun=4.21\,\,\,\,$\alpha = 1.5$ }\par\medskip
\includegraphics[trim=20 20 0 75,clip,width=0.8\textwidth]{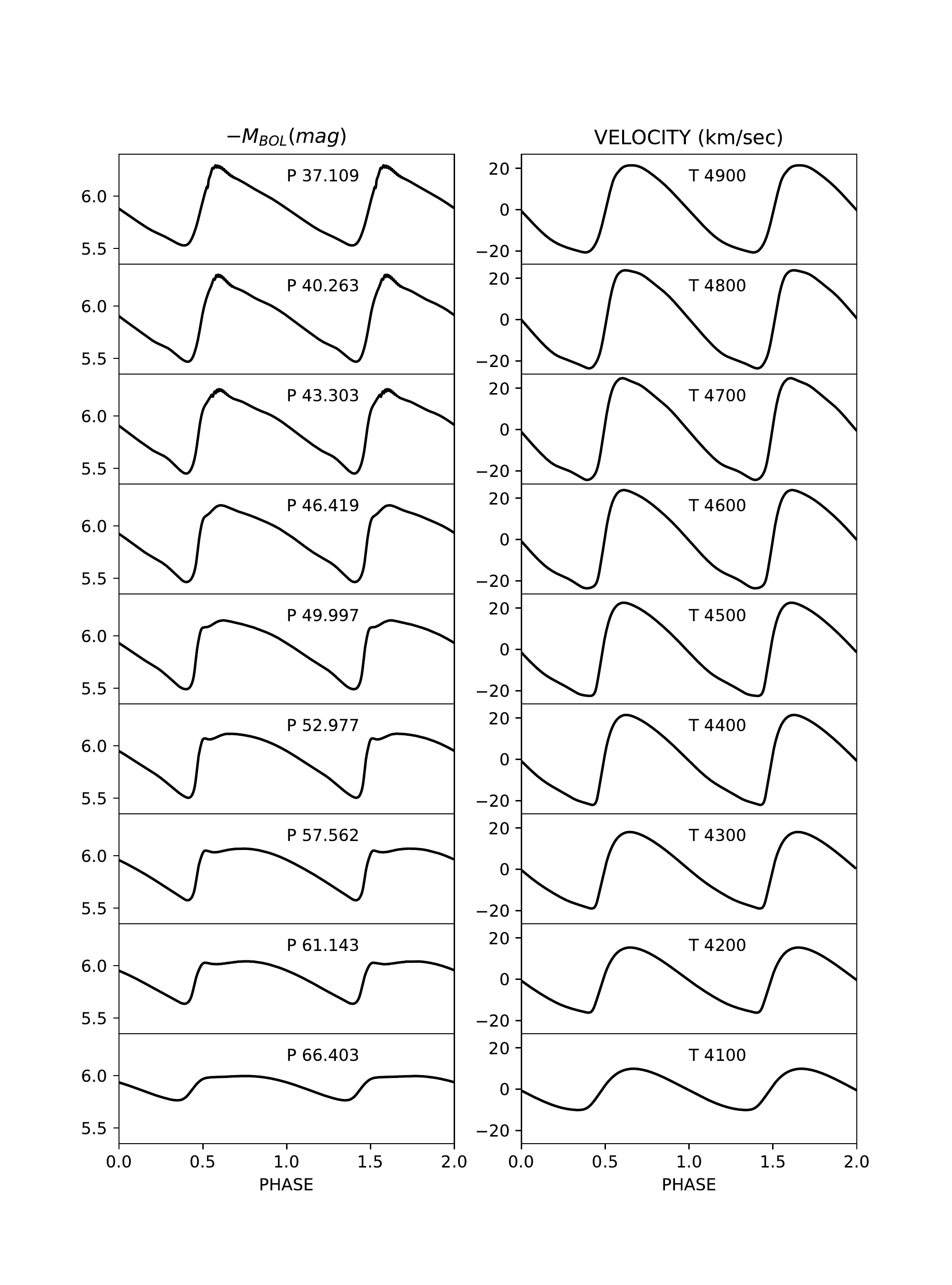}
\par FIG.3-Continued.
\label{Fig:lc_vr_4.21_11.0_1.5_ca}
\end{figure}
\clearpage

\subsubsection{The effect of the assumed Mass-Luminosity relation}
To show the effect of a variation of the ML relation on the predicted
light and radial velocity curves,we chose two models whose trend is representative of all other models.\\ In the panels (a) e (b) of Figure (\ref{fig:effect_ml_f}) we show the light and radial velocity curves of a
F model respectively, at fixed $\frac{M}{M_{\odot}}=9.0$, $ T_{eff} = 4600 K$ and
$\alpha=1.5$ but for three different levels of luminosity (case A, B
and C). As in Figure (\ref{fig:effect_ml_f}), the panels (a) and (b) in Figure (\ref{fig:effect_ml_fo}) show the similar comparison for a FO model with
$\frac{M}{M_{\odot}}=3.0$, $ T_{eff} = 6300 K$ and $\alpha=1.5$.
These plots confirm that both the morphology and amplitude of light
and radial velocity curves depend on the assumed ML relation.

Moreover, inspection of  B and C model sets indicates that in these cases
longer periods are found at fixed mass due to the increased luminosity
levels.  As a consequence, ''bump Cepheids'' are found at lower masses and the center of the HP is found at slightly shorter periods. This trend, once the metallicity is known,  makes the HP phenomenon a useful tracer of Cepheid ML relation.

\subsubsection{The effect of the assumed superadiabatic convective efficiency}
As discussed above, the main effect of superadiabatic convection is to quench pulsation so that lower pulsation amplitudes are expected as $\alpha$ increases from 1.5 to 1.7 and 1.9. Assuming a canonical ML, Figure \ref{fig:pyramid_alfa_f} and Figure \ref{fig:pyramid_alfa_fo} show the comparison between the light and radial velocity curves obtained for the three values of $\alpha$ but at fixed stellar parameters correspond to a F and a FO model with $\frac{M}{M_{\odot}}=4.0$, $ T_{eff} = 5900 K$ and and $\frac{M}{M_{\odot}}=4.0$, $ T_{eff} = 6200 K$, respectively. The morphology of the curves gets smoother and the pulsation amplitude decreases as $\alpha$ increases. In particular the FO pulsation disappears for $\alpha=1.9$ because at this value of $\alpha$ the quenching effect is very efficient and the pulsation disappears. The same trend is followed by all other models.

\subsection{The Period-Radius and the Period-Mass-Radius relations}
\begin{figure}
\centering
\includegraphics[width=0.6\textwidth]{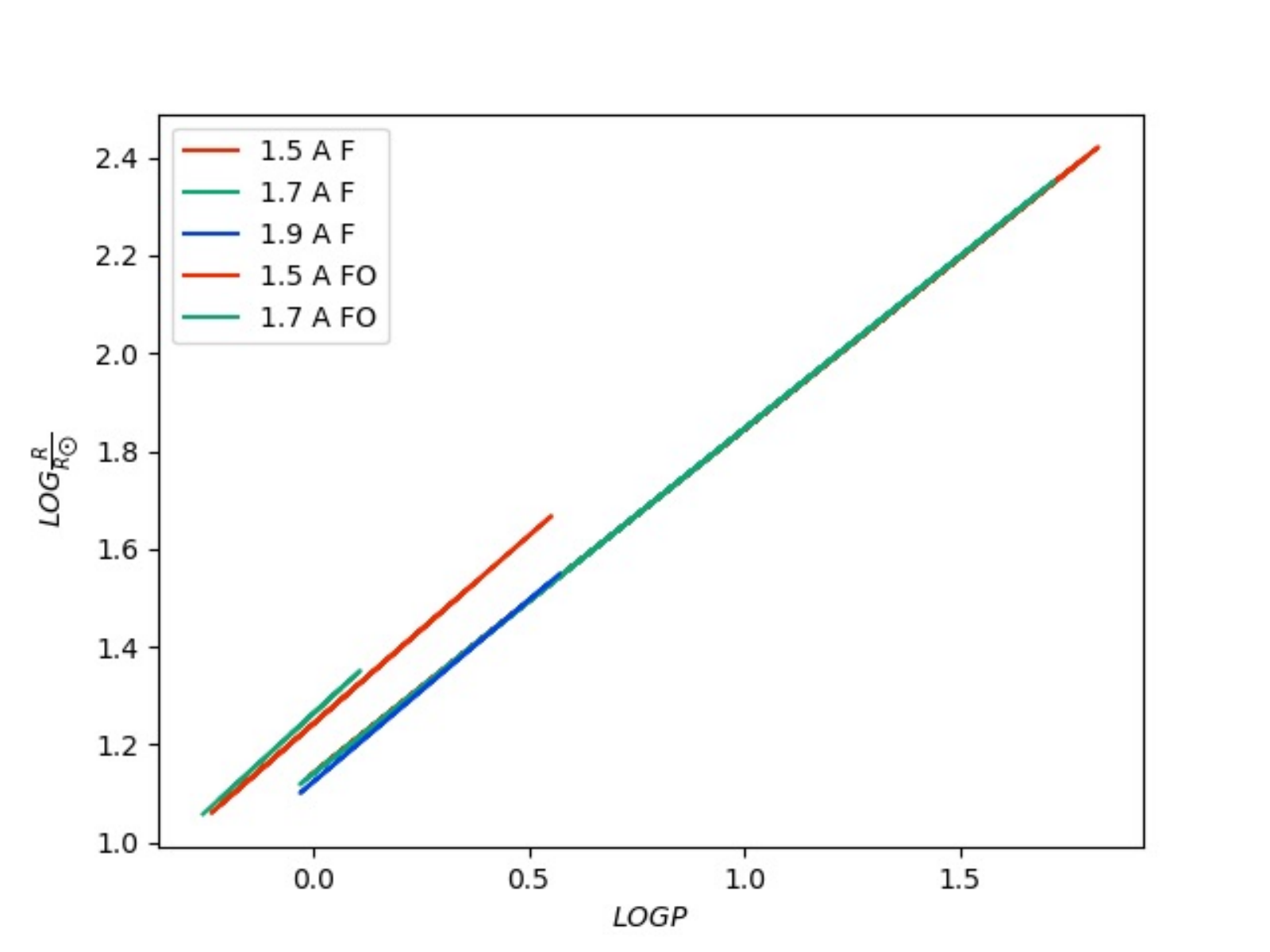}
\caption{PR relations for F and FO Galactic Cepheids derived by adopting canonical ML relation and $\alpha = 1.5$, $\alpha = 1.7$, $\alpha = 1.9$.}
\label{Fig:pr_alfa}
\end{figure}

\begin{figure}
\centering
\includegraphics[trim=5 0 30 0, angle = 0, width=0.9\textwidth]{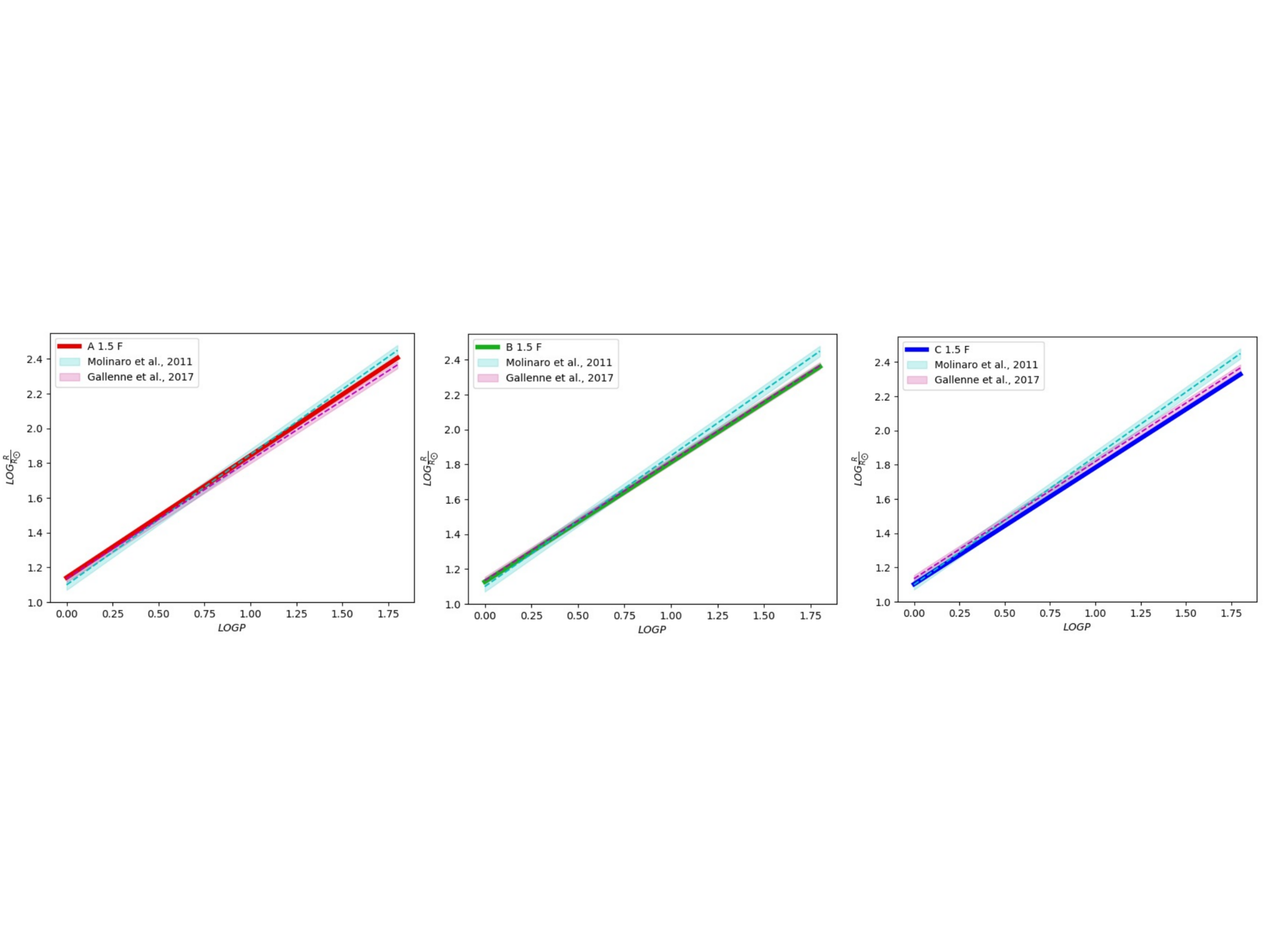}
\caption{PR relations for F Galactic Cepheids derived by adopting $\alpha = 1.5$ and A, B , C ML relations compared with similar results available in literature.}
\label{fig:effect_3ml}
\end{figure}

An important aspect of Cepheid research concerns the use of CC to infer stellar radii. CC are known to obey to both PR and PMR relations, the former involving an averaging operation over the finite width in effective temperature of the instability strip \citep{bgm2001}.
The PR relations are used to derive the stellar radii directly from the pulsation periods, whereas PMR relations can be used to infer an independent value of the stellar mass, once known the period and the radius, to be compared with evolutionary mass estimates. The coefficients of the PR and PMR relations derived from current nonlinear model sets are reported in Tables \ref{pr_f_fo} and \ref{pmr_f_fo}  for the F and FO models, respectively.
Figure \ref{Fig:pr_alfa} shows the PR relations assuming the canonical ML relation for the three values of $\alpha$ for F and FO models and Figure \ref{fig:effect_3ml} shows the PR relations at fixed $\alpha = 1.5$ for ML relation from case A to B and C for F models.
We confirm previous results by \citet{bcm1998} that the PR relation does not vary considerably with the different assumptions of the ML relation (see Table \ref{pr_f_fo}). Moreover varying the efficiency of superadiabatic convection has a mild effect on the PR coefficients.

Furthermore, we perform a comparison with literature relations of our PR (Figure \ref{fig:effect_3ml}). We confirm a general good agreement with \citet{mrm2012} and \citet{gal2017} PR relations. However we better reproduce \citet{mrm2012} relation at shorter periods whereas the opposite occurs with \citet{gal2017} relation. We finally note that the PR relations obtained by \citet{mrm2012} and \citet{gal2017} depend
on the assumed projection-factor (p-factor) value\footnote{The p-factor is the parameter which connects the observed radial velocity to the model radial velocity (see e.g. \citet{gal2017})}, while those derived by
the models do not depend on this parameter \citep{ragosta2019} . Therefore a comparison between 
these two independent derivations allows us to put constraints on the value
of the p-factor (e.g.\citet{natale2008,mp2013}) that plays a key role in the study of pulsating
stars. Even if the investigation of the p-factor and of its dependence on the pulsation period is beyond the scope of the present paper, from the combination of our extended atlas of pulsation
models with the large sample of
radial velocity curves that will be provided by the \textsl{Gaia} mission, we will be able to constrain the p-factor with a much more robust statistics than in previous studies.

\begin{figure*}
\centering
\gridline{\fig{{light_curve_4600_9.0_1.5_}.pdf}{0.5\textwidth}{(a)}
          \fig{{velocity_curve_4600_9.0_1.5_}.pdf}{0.5\textwidth}{(b)}
          }
\caption{Light (a) and radial velocity (b) curves of $ T_{eff} = 4600 K$, \msun=9.0, $\alpha=1.5$ F model for three levels of luminosity.}
\label{fig:effect_ml_f}
\end{figure*}

\begin{figure*}
\centering
\gridline{\fig{{light_curve_6300_3.0_1.5_fofo}.pdf}{0.5\textwidth}{(a)}
          \fig{{velocity_curve_6300_3.0_1.5_fo}.pdf}{0.5\textwidth}{(b)}
          }
\caption{Light (a) and radial velocity (b) curves of $ T_{eff} = 6300 K$, \msun=3.0, $\alpha=1.5$ FO model for different levels of luminosity.}
\label{fig:effect_ml_fo}
\end{figure*}

\begin{figure*}
\centering
\gridline{\fig{{light_curve_5900_2.74_4.0_ca}.pdf}{0.5\textwidth}{(a)}
          \fig{{velocity_curve_5900_2.74_4.0_ca}.pdf}{0.5\textwidth}{(b)}
          }

\caption{Light (a) and radial velocity (b) curves for \msun=4.0, $ T_{eff} = 5900 K$ F model for the canonical ML assumption.}
\label{fig:pyramid_alfa_f}
\end{figure*}

\begin{figure*}
\centering
\gridline{\fig{{curves_smooth}.pdf}{0.5\textwidth}{(a)}
          \fig{{velocity_curve_6200_2.74_4.0_ca}.pdf}{0.5\textwidth}{(b)}
          }

\caption{Light (a) and radial velocity (b) curves for \msun=4.0, $ T_{eff} = 6200 K$ FO model for the canonical ML assumption.}
\label{fig:pyramid_alfa_fo}
\end{figure*}
\clearpage

\begin{ThreePartTable}
\begin{longtable}{ccccccc}
\caption{\label{pr_f_fo} PR coefficients ($\log R=a+b\log P$) for F and FO Galactic Cepheids derived by adopting A, B , C ML relations.}\\
\hline\hline
$\alpha$ &ML&a&b& $\sigma_{a}$& $\sigma_{b}$&$R^2$\\
\hline
F\\
\hline
\endfirsthead
\caption{continued.}\\
\hline\hline
$\alpha$ &ML&a&b& $\sigma_{a}$& $\sigma_{b}$&$R^2$\\
\hline
\endhead
\hline
\endfoot
1.5&A&1.142&0.702&0.004&0.003&0.998\\
1.5&B&1.128&0.685&0.005&0.003&0.998\\
1.5&C&1.104&0.680&0.005&0.003&0.998\\
1.7&A&1.140&0.705&0.004&0.003&0.999\\
1.7&B&1.126&0.685&0.005&0.003&0.999\\
1.7&C&1.105&0.678&0.005&0.003&0.999\\
1.9&A&1.124&0.743&0.003&0.007&0.999\\
1.9&B&1.101&0.729&0.003&0.008&0.999\\
1.9&C&1.077&0.715&0.003&0.005&0.999\\
\hline
FO\\
\hline
1.5&A&1.242&0.768&0.001&0.005&0.999\\
1.5&B&1.216&0.762&0.003&0.015&0.997\\
1.5&C&1.193&0.742&0.003&0.009&0.997\\
1.7&A&1.243&0.773&0.002&0.009&0.840\\
\hline
\end{longtable}
\end{ThreePartTable}

\begin{ThreePartTable}
\begin{longtable}{ccccccccc}
\caption{\label{pmr_f_fo} PMR coefficients ($\log R=a+b\log P+c\log M$) for F and FO Galactic Cepheids derived by adopting A, B , C ML relations.}\\
\hline\hline
$\alpha$&ML&a&b&c&$\sigma_{a}$&$\sigma_{b}$&$\sigma_{c}$&$R^2$\\
\hline
F\\
\hline
\endfirsthead
\caption{continued.}\\
\hline\hline
$\alpha$&ML&a&b&c&$\sigma_{a}$&$\sigma_{b}$&$\sigma_{c}$&$R^2$\\
\hline
\endhead
\hline
\endfoot
1.5&A&-1.641&-0.890&1.830&0.007&0.06&0.03&0.999\\
1.5&B&-1.709&-0.920&1.874&0.01&0.072&0.03&0.998\\
1.5&C&-1.721&-0.687&1.784&0.01&0.06&0.03&0.998\\
1.7&A&-1.642&-1.144&1.948&0.008&0.14&0.06&0.999\\
1.7&B&-1.725&-1.194&2.001&0.01&0.13&0.06&0.999\\
1.7&C&-1.687&-0.583&1.728&0.02&0.11&0.05&0.999\\
1.9&A&-1.570&-0.778&1.737&0.02&0.3&0.1&0.999\\
1.9&B&-1.573&-0.720&1.709&0.01&0.09&0.05&0.999\\
1.9&C&-1.587&-0.547&1.654&0.01&0.06&0.03&0.999\\
\hline
FO\\
\hline
1.5&A&-1.659&-0.564&1.590&0.005&0.06&0.03&0.999\\
1.5&B&-1.695&-0.779&1.707&0.02&0.1007&0.05&0.999\\
1.5&C&-1.738&-0.698&1.704&0.01&0.06&0.03&0.999\\
1.7&A&-1.644&-0.589&1.591&0.01&0.1&0.05&0.902\\
\hline
\end{longtable}
\end{ThreePartTable}

\section{Predicted light curves, mean magnitudes and colors in the \textsl{Gaia} photometric system}
\begin{figure*}
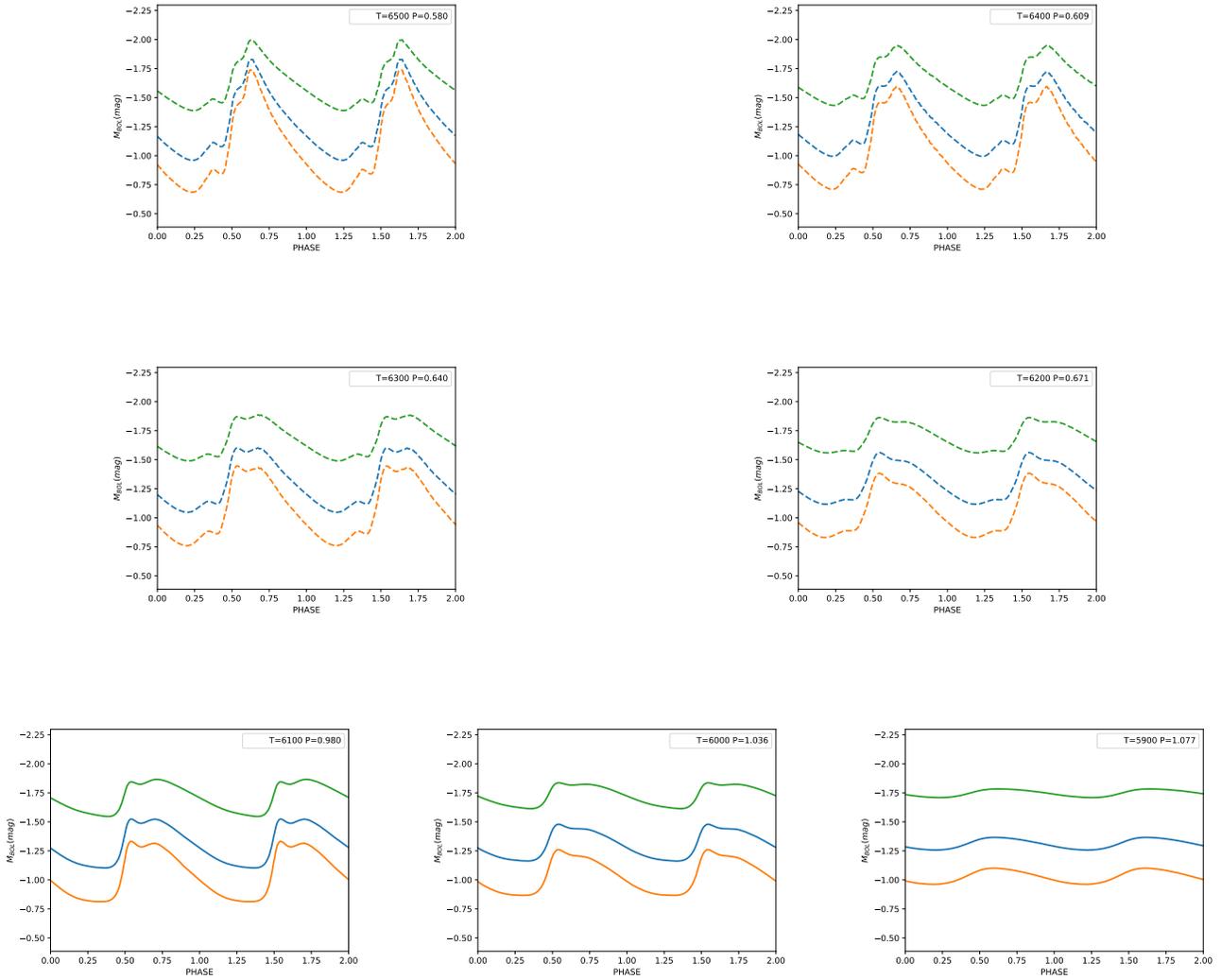

\centering
\textbf{\msun=3.0\,\,\,\,\lsun=2.32\,\,\,\,$\alpha = 1.5$ }\par\medskip
\gridline{\fig{{lg_split_tratt_n1_6500_2.32_3.0_1.5_ca}.pdf}{0.30\textwidth}{}
          \fig{{lg_split_tratt_n1_6400_2.32_3.0_1.5_ca}.pdf}{0.30\textwidth}{}
          }
\gridline{\fig{{lg_split_tratt_n1_6300_2.32_3.0_1.5_ca}.pdf}{0.30\textwidth}{}
          \fig{{lg_split_tratt_n1_6200_2.32_3.0_1.5_ca}.pdf}{0.30\textwidth}{}
          }
\gridline{\fig{{lg_split_n0_6100_2.32_3.0_1.5_ca}.pdf}{0.30\textwidth}{} 
          \fig{{lg_split_n0_6000_2.32_3.0_1.5_ca}.pdf}{0.30\textwidth}{}
          \fig{{lg_split_n0_5900_2.32_3.0_1.5_ca}.pdf}{0.30\textwidth}{}
          }

\caption{ Theoretical \textsl{Gaia} light curves for a sequence of non linear F (solid line) and FO models (dashed line) derived at fixed mass, luminosity, $\alpha$ parameter (see labeled values on the top of the plot) adopting the canonical ML relation.}
\label{fig:gaia_3.0}
\end{figure*}

The bolometric light curves presented in Section 3 have been converted in the \textsl{Gaia} photometric system passbands, namely $G$, $G_{BP}$ and $G_{RP}$, using the ATLAS9 non-overshooting model atmospheres \citep{castelli2003}. This provides the first theoretical catalogue of \textsl{Gaia} light curves.
The predicted light curves for $\alpha = 1.5$ and masses ranging from 3 to 11 $M_{\odot}$ are shown in Figure  \ref{fig:gaia_3.0} where the green line indicate the $G_{BP}$ band, the blue line the $G$ band and the orange line the $G_{RP}$ band. Dashed and solid lines represent the FO and F models, respectively. On each light curve the effective temperature in kelvin and the period in days of the model is labeled. The complete atlas of the light curves in the \textsl{Gaia} photometric system for the various assumptions about the ML relation and the superadiabatic convection efficiency are available in the Appendix. We note that the morphology
of the predicted \textsl{Gaia} light curves follow the features of the bolometric ones.
The converted light curves allow us to derive intensity-averaged mean magnitudes and colors in the \textsl{Gaia} filters, namely magnitudes $<M_{G}>$,  $<M_{G_{BP}}>$, $<M_{G_{RP}}>$ and color $<G_{BP}>$ - $<G_{RP}>$. The three mean magnitudes are reported in Tables \ref{gaia_mean_f} and \ref{gaia_mean_fo} for each F and FO model.
\clearpage

\clearpage

\begin{figure*}
\centering
\textbf{\msun=4.0\,\,\,\,\lsun=2.74\,\,\,\,$\alpha = 1.5$ }\par\medskip
\gridline{\fig{{lg_split_tratt_n1_6400_2.74_4.0_1.5_ca}.pdf}{0.30\textwidth}{}
          \fig{{lg_split_tratt_n1_6300_2.74_4.0_1.5_ca}.pdf}{0.30\textwidth}{}
          \fig{{lg_split_tratt_n1_6200_2.74_4.0_1.5_ca}.pdf}{0.30\textwidth}{}
          }
\gridline{\fig{{lg_split_tratt_n1_6100_2.74_4.0_1.5_ca}.pdf}{0.30\textwidth}{}
          \fig{{lg_split_tratt_n1_6000_2.74_4.0_1.5_ca}.pdf}{0.30\textwidth}{}
          \fig{{lg_split_tratt_n1_5900_2.74_4.0_1.5_ca}.pdf}{0.30\textwidth}{}
          }
\gridline{\fig{{lg_split_n0_5900_2.74_4.0_1.5_ca}.pdf}{0.30\textwidth}{} 
          \fig{{lg_split_n0_5800_2.74_4.0_1.5_ca}.pdf}{0.30\textwidth}{}
          \fig{{lg_split_n0_5700_2.74_4.0_1.5_ca}.pdf}{0.30\textwidth}{)}
          }
\gridline{\fig{{lg_split_n0_5600_2.74_4.0_1.5_ca}.pdf}{0.30\textwidth}{} 
          \fig{{lg_split_n0_5500_2.74_4.0_1.5_ca}.pdf}{0.30\textwidth}{}
          }
\par{FIG.10-Continued.}
\label{fig:gaia_4.0}
\end{figure*}
\clearpage

\begin{figure*}
\centering
\textbf{\msun=5.0\,\,\,\,\lsun=3.07\,\,\,\,$\alpha = 1.5$ }\par\medskip
\gridline{\fig{{lg_split_tratt_n1_6100_3.07_5.0_1.5_ca}.pdf}{0.30\textwidth}{}
          \fig{{lg_split_tratt_n1_6000_3.07_5.0_1.5_ca}.pdf}{0.30\textwidth}{}
          }
\gridline{\fig{{lg_split_tratt_n1_5900_3.07_5.0_1.5_ca}.pdf}{0.30\textwidth}{}
          \fig{{lg_split_tratt_n1_5800_3.07_5.0_1.5_ca}.pdf}{0.30\textwidth}{}
          }
\gridline{\fig{{lg_split_n0_5800_3.07_5.0_1.5_ca}.pdf}{0.30\textwidth}{} 
          \fig{{lg_split_n0_5700_3.07_5.0_1.5_ca}.pdf}{0.30\textwidth}{}
          \fig{{lg_split_n0_5600_3.07_5.0_1.5_ca}.pdf}{0.30\textwidth}{}
          }
\gridline{\fig{{lg_split_n0_5500_3.07_5.0_1.5_ca}.pdf}{0.30\textwidth}{} 
          \fig{{lg_split_n0_5400_3.07_5.0_1.5_ca}.pdf}{0.30\textwidth}{}
          \fig{{lg_split_n0_5300_3.07_5.0_1.5_ca}.pdf}{0.30\textwidth}{}
          }
\par{FIG.10-Continued.}
\label{fig:gaia_5.0}
\end{figure*}
\clearpage

\begin{figure*}
\centering
\textbf{\msun=6.0\,\,\,\,\lsun=3.33\,\,\,\,$\alpha = 1.5$ }\par\medskip
\gridline{\fig{{lg_split_tratt_n1_5800_3.33_6.0_1.5_ca}.pdf}{0.30\textwidth}{}
          }
\gridline{\fig{{lg_split_n0_5800_3.33_6.0_1.5_ca}.pdf}{0.30\textwidth}{}
          \fig{{lg_split_n0_5700_3.33_6.0_1.5_ca}.pdf}{0.30\textwidth}{}
          \fig{{lg_split_n0_5600_3.33_6.0_1.5_ca}.pdf}{0.30\textwidth}{}
          }
\gridline{\fig{{lg_split_n0_5500_3.33_6.0_1.5_ca}.pdf}{0.30\textwidth}{} 
          \fig{{lg_split_n0_5400_3.33_6.0_1.5_ca}.pdf}{0.30\textwidth}{}
          \fig{{lg_split_n0_5300_3.33_6.0_1.5_ca}.pdf}{0.30\textwidth}{}
          }
\gridline{\fig{{lg_split_n0_5200_3.33_6.0_1.5_ca}.pdf}{0.30\textwidth}{} 
          \fig{{lg_split_n0_5100_3.33_6.0_1.5_ca}.pdf}{0.30\textwidth}{}
          \fig{{lg_split_n0_5000_3.33_6.0_1.5_ca}.pdf}{0.30\textwidth}{}
          }
\par{FIG.10-Continued.}
\label{fig:gaia_6.0}
\end{figure*}
\clearpage

\begin{figure*}
\centering
\textbf{\msun=7.0\,\,\,\,\lsun=3.56\,\,\,\,$\alpha = 1.5$ }\par\medskip
\gridline{\fig{{lg_split_n0_5500_3.56_7.0_1.5_ca}.pdf}{0.30\textwidth}{}
          \fig{{lg_split_n0_5400_3.56_7.0_1.5_ca}.pdf}{0.30\textwidth}{}
          }
\gridline{\fig{{lg_split_n0_5300_3.56_7.0_1.5_ca}.pdf}{0.30\textwidth}{}
          \fig{{lg_split_n0_5200_3.56_7.0_1.5_ca}.pdf}{0.30\textwidth}{}
          \fig{{lg_split_n0_5100_3.56_7.0_1.5_ca}.pdf}{0.30\textwidth}{}
          }
\gridline{\fig{{lg_split_n0_5000_3.56_7.0_1.5_ca}.pdf}{0.30\textwidth}{} 
          \fig{{lg_split_n0_4900_3.56_7.0_1.5_ca}.pdf}{0.30\textwidth}{}
          \fig{{lg_split_n0_4800_3.56_7.0_1.5_ca}.pdf}{0.30\textwidth}{}
          }

\par{FIG.10-Continued.}
\label{fig:gaia_7.0}
\end{figure*}
\clearpage

\begin{figure*}
\centering
\textbf{\msun=8.0\,\,\,\,\lsun=3.75\,\,\,\,$\alpha = 1.5$ }\par\medskip
\gridline{\fig{{lg_split_n0_5400_3.75_8.0_1.5_ca}.pdf}{0.30\textwidth}{}
          \fig{{lg_split_n0_5300_3.75_8.0_1.5_ca}.pdf}{0.30\textwidth}{}
          \fig{{lg_split_n0_5200_3.75_8.0_1.5_ca}.pdf}{0.30\textwidth}{}
          }
\gridline{\fig{{lg_split_n0_5100_3.75_8.0_1.5_ca}.pdf}{0.30\textwidth}{}
          \fig{{lg_split_n0_5000_3.75_8.0_1.5_ca}.pdf}{0.30\textwidth}{}
          \fig{{lg_split_n0_4900_3.75_8.0_1.5_ca}.pdf}{0.30\textwidth}{}
          }
\gridline{\fig{{lg_split_n0_4800_3.75_8.0_1.5_ca}.pdf}{0.30\textwidth}{} 
          \fig{{lg_split_n0_4700_3.75_8.0_1.5_ca}.pdf}{0.30\textwidth}{}
          \fig{{lg_split_n0_4600_3.75_8.0_1.5_ca}.pdf}{0.30\textwidth}{}
          }

\par{FIG.10-Continued.}
\label{fig:gaia_8.0}
\end{figure*}
\clearpage

\begin{figure*}
\centering
\textbf{\msun=9.0\,\,\,\,\lsun=3.92\,\,\,\,$\alpha = 1.5$ }\par\medskip
\gridline{\fig{{lg_split_n0_5200_3.92_9.0_1.5_ca}.pdf}{0.30\textwidth}{}
          \fig{{lg_split_n0_5100_3.92_9.0_1.5_ca}.pdf}{0.30\textwidth}{}
          \fig{{lg_split_n0_5000_3.92_9.0_1.5_ca}.pdf}{0.30\textwidth}{}
          }
\gridline{\fig{{lg_split_n0_4900_3.92_9.0_1.5_ca}.pdf}{0.30\textwidth}{}
          \fig{{lg_split_n0_4800_3.92_9.0_1.5_ca}.pdf}{0.30\textwidth}{}
          \fig{{lg_split_n0_4700_3.92_9.0_1.5_ca}.pdf}{0.30\textwidth}{}
          }
\gridline{\fig{{lg_split_n0_4600_3.92_9.0_1.5_ca}.pdf}{0.30\textwidth}{} 
          \fig{{lg_split_n0_4500_3.92_9.0_1.5_ca}.pdf}{0.30\textwidth}{}
          \fig{{lg_split_n0_4400_3.92_9.0_1.5_ca}.pdf}{0.30\textwidth}{}
          }

\par{FIG.10-Continued.}
\label{fig:gaia_9.0}
\end{figure*}
\clearpage

\begin{figure*}
\centering
\textbf{\msun=10.0\,\,\,\,\lsun=4.08\,\,\,\,$\alpha = 1.5$ }\par\medskip
\gridline{\fig{{lg_split_n0_5100_4.08_10.0_1.5_ca}.pdf}{0.30\textwidth}{}
          \fig{{lg_split_n0_5000_4.08_10.0_1.5_ca}.pdf}{0.30\textwidth}{}
          }
\gridline{\fig{{lg_split_n0_4900_4.08_10.0_1.5_ca}.pdf}{0.30\textwidth}{}
          \fig{{lg_split_n0_4800_4.08_10.0_1.5_ca}.pdf}{0.30\textwidth}{}
          }
\gridline{\fig{{lg_split_n0_4700_4.08_10.0_1.5_ca}.pdf}{0.30\textwidth}{} 
          \fig{{lg_split_n0_4600_4.08_10.0_1.5_ca}.pdf}{0.30\textwidth}{}
          \fig{{lg_split_n0_4500_4.08_10.0_1.5_ca}.pdf}{0.30\textwidth}{}
          }
\gridline{\fig{{lg_split_n0_4400_4.08_10.0_1.5_ca}.pdf}{0.30\textwidth}{} 
          \fig{{lg_split_n0_4300_4.08_10.0_1.5_ca}.pdf}{0.30\textwidth}{}
          \fig{{lg_split_n0_4200_4.08_10.0_1.5_ca}.pdf}{0.30\textwidth}{}
          }

\par{FIG.10-Continued.}
\label{fig:gaia_10.0}
\end{figure*}
\clearpage

\begin{figure*}
\centering
\textbf{\msun=11.0\,\,\,\,\lsun=4.21\,\,\,\,$\alpha = 1.5$ }\par\medskip
\gridline{\fig{{lg_split_n0_4900_4.21_11.0_1.5_ca}.pdf}{0.30\textwidth}{}
          \fig{{lg_split_n0_4800_4.21_11.0_1.5_ca}.pdf}{0.30\textwidth}{}
          \fig{{lg_split_n0_4700_4.21_11.0_1.5_ca}.pdf}{0.30\textwidth}{}
          }
\gridline{\fig{{lg_split_n0_4600_4.21_11.0_1.5_ca}.pdf}{0.30\textwidth}{}
          \fig{{lg_split_n0_4500_4.21_11.0_1.5_ca}.pdf}{0.30\textwidth}{}
          \fig{{lg_split_n0_4400_4.21_11.0_1.5_ca}.pdf}{0.30\textwidth}{}
          }
\gridline{\fig{{lg_split_n0_4300_4.21_11.0_1.5_ca}.pdf}{0.30\textwidth}{}
          \fig{{lg_split_n0_4200_4.21_11.0_1.5_ca}.pdf}{0.30\textwidth}{}
          \fig{{lg_split_n0_4100_4.21_11.0_1.5_ca}.pdf}{0.30\textwidth}{}
          }          

\par{FIG.10-Continued.}
\label{fig:gaia_11.0}
\end{figure*}
\clearpage

\begin{ThreePartTable}
\begin{TableNotes}
\footnotesize 
\item[a] Stellar mass (solar units).
\item[b] Logarithmic luminosity (solar units).
\item[c] Effective temperature(K).
\item[d] Mixing length parameter.
\item[e] Mass-Luminosity relation.
\item[f] \textsl{Gaia} passband G.
\item[f] \textsl{Gaia} passband $G_{BP}$.
\item[h] \textsl{Gaia} passband $G_{RP}$.
\end{TableNotes}
\begin{longtable}{cccccccccc}
\caption{\label{gaia_mean_f} Mean magnitudes in the \textsl{Gaia} filters for F models at varying the ML relation and the $\alpha$ pareameter. Full tables are available in the Appendix.}\\
\hline\hline
&&&Z=0.02 & Y= 0.28\\
M\tnote{a}&logL\tnote{b}&$T_{eff}$\tnote{c}&$\alpha$\tnote{d}&ML\tnote{e}&G\tnote{f}& $G_{BP}$\tnote{g}&  $G_{RP}$\tnote{h}\\
\hline
(1)&(2)&(3)&(4)&(5)&(6)&(7)&(8)\\
\hline
\endfirsthead
\caption{continued.}\\
\hline\hline
M\tnote{a}&logL\tnote{b}&$T_{eff}$\tnote{c}&$\alpha$\tnote{d}&ML\tnote{e}&G\tnote{f}& $G_{BP}$\tnote{g}& $G_{RP}$\tnote{h}\\
\hline
(1)&(2)&(3)&(4)&(5)&(6)&(7)&(8)\\
\hline
\endhead
\hline
\endfoot
3.0&2.32&5900&1.5&A&1.31&-1.03&-1.75\\
3.0&2.32&6000&1.5&A&1.31&-1.05&-1.73\\
...\\
4.0&2.74&5500&1.5&A&2.34&-1.99&-2.85\\
4.0&2.74&5600&1.5&A&2.34&-2.01&-2.83\\
...\\
5.0&3.07&5300&1.5&A&3.13&-2.74&-3.68\\
5.0&3.07&5400&1.5&A&3.14&-2.77&-3.67\\
...\\
6.0&3.33&5000&1.5&A&3.75&-3.30&-4.36\\
6.0&3.33&5100&1.5&A&3.76&-3.33&-4.36\\
...\\
7.0&3.56&4800&1.5&A&4.26&-3.77&-4.91\\
7.0&3.56&4900&1.5&A&4.28&-3.81&-4.90\\
...\\
8.0&3.75&4600&1.5&A&4.70&-4.16&-5.39\\
8.0&3.75&4700&1.5&A&4.72&-4.21&-5.39\\
...\\
9.0&3.92&4400&1.5&A&5.07&-4.48&-5.79\\
9.0&3.92&4500&1.5&A&5.09&-4.53&-5.80\\
...\\
10.0&4.08&4200&1.5&A&5.38&-4.75&-6.15\\
10.0&4.08&4300&1.5&A&5.41&-4.80&-6.16\\
...\\
11.0&4.21&4100&1.5&A&5.68&-5.02&-6.47\\
11.0&4.21&4200&1.5&A&5.72&-5.08&-6.49\\
...\\
\hline
\insertTableNotes  
\end{longtable}
\end{ThreePartTable}
\clearpage

\begin{ThreePartTable}
\begin{TableNotes}
\footnotesize 
\item[a] Stellar mass (solar units).
\item[b] Logarithmic luminosity (solar units).
\item[c] Effective temperature(K).
\item[d] Mixing length parameter.
\item[e] Mass-Luminosity relation.
\item[f] \textsl{Gaia} passband G.
\item[g] \textsl{Gaia} passband $G_{BP}$.
\item[h] \textsl{Gaia} passband $G_{RP}$.
\end{TableNotes}
\begin{longtable}{cccccccccc}
\caption{\label{gaia_mean_fo} The same as in Table \ref{gaia_mean_f} but for FO models.}\\
\hline\hline
&&&Z=0.02 & Y= 0.28\\
M\tnote{a}&logL\tnote{b}&$T_{eff}$\tnote{c}&$\alpha$\tnote{d}&ML\tnote{e}&G\tnote{f}& $G_{BP}$\tnote{g}&  $G_{RP}$\tnote{h}\\
\hline
(1)&(2)&(3)&(4)&(5)&(6)&(7)&(8)\\
\hline
\endfirsthead
\caption{continued.}\\
\hline\hline
M\tnote{a}&logL\tnote{b}&$T_{eff}$\tnote{c}&$\alpha$\tnote{d}&ML\tnote{e}&G\tnote{f}& $G_{BP}$\tnote{g}&  $G_{RP}$\tnote{h}\\
\hline
(1)&(2)&(3)&(4)&(5)&(6)&(7)&(8)\\
\hline
\endhead
\hline
\endfoot
3.0&2.32&6200&1.5&A&-1.32&-1.08&-1.70\\
3.0&2.32&6300&1.5&A&-1.32&-1.10&-1.68\\
...\\
4.0&2.74&5900&1.5&A&-2.36&-2.08&-2.80\\
4.0&2.74&6000&1.5&A&-2.37&-2.10&-2.78\\
...\\
5.0&3.07&5800&1.5&A&-3.17&-2.88&-3.63\\
5.0&3.07&5900&1.5&A&-3.18&-2.90&-3.61\\
...\\
6.0&3.33&5800&1.5&A&-3.84&-3.54&-4.29\\
\hline
\insertTableNotes  
\end{longtable}
\end{ThreePartTable}
\clearpage

\subsection{The Period-Luminosity-Color and the Period-Wesenheit relations in the \textsl{Gaia} filters}
The mean magnitudes and colors derived in the previous section can be
used to derive the first theoretical PLC and PW relations in the \textsl{Gaia}
filters. The coefficients of these relations at varying the ML relation and the efficiency of the superadiabatic convection, are reported in Table \ref{plc_f_fo} and Table \ref{w_f_fo} for F and FO models, respectively. To derive the Wesenheit magnitude we adopt the relation provided by \citet{rip2019} $<W>=<G> -1.9 <G_{BP}-G_{RP}>$. Both the PLC and the PW relations hold for each individual pulsator thus allowing us to derive
individual distances of observed Cepheids in the \textsl{Gaia} database. We
notice that the PLC and PW relations, and in turn the individual
distances derived by applying them to the observed
pulsators, depend on the assumed ML relation but are almost
insensitive to the value of the mixing length parameter. In particular assuming $\alpha=1.5$, if
we consider a F mode Cepheid
pulsator with $P=10$ days and $<G_{BP}>$ - $<G_{RP}>$ = 1.0 mag, the
 $<G>$ magnitude obtained from the theoretical PLC relation varies from $<G>$= $-4.11$ mag at the canonical ML (case A) to $<G>$= $-3.94$ mag for a ML brighter by 0.2 dex (case B)  and to $<G>$ = $-3.79$ mag
for a ML brighter by 0.4 dex (case C). Consequently assuming $P=10$ days and $<G_{BP}>$ - $<G_{RP}>$ = 1.0 mag,
the difference in the predicted $<G>$ magnitude amounts to $0.1$ mag
and $0.3$ mag
when the ML relation changes from case A to B and C, respectively and assuming the canonical case A, the $<G> $ magnitude obtained from the PLC relation can change up to 0.1 mag when moving from $\alpha=1.5$ to $\alpha=1.9$. In order to better exemplify what could occur in typical extragalactic distance scale applications, we performed the same kind of test with F mode PLC relations at fixed periods of 30 and 100 days and $<G_{BP}>$ - $<G_{RP}>$ = 1.0 mag. As a result, we found that, assuming $\alpha=1.5$ and  $P=30$ days, $<G>$ varies from $-5.91$ (case A) to $-5.73$ (case B) and  $-5.57$  case (C), whereas for a still longer period  (P=100 days) $<G>$  varies from $-7.89$  (case A) to  $-7.7$  (case B) and $-7.51$ (case C). On this basis we conclude that the effects related to superadiabatic convection on the predicted $<G>$ magnitude amount to about 0.15 mag and 0.20 mag for  $P=30$ days and $P=100$ days respectively, whereas the effects related to variations in the ML relation can be as large as 0.4 mag, for the two period assumptions. 
Similar considerations hold for the predicted F mode PW relations. 
As for FO pulsators, both PLC and PW relations are
insensitive to variations in the efficiency of superadiabatic convection.”

\begin{ThreePartTable}
\begin{longtable}{cccccccccc}
\caption{\label{plc_f_fo} PLC coefficients ($<G>$=a+b$\log P$ +c($<G_{BP}>$ - $<G_{RP}>$) for F and FO Galactic Cepheids derived by adopting A, B , C ML relations and $\alpha = 1.5$, $\alpha = 1.7$ and $\alpha = 1.9$ in the \textsl{Gaia} filters.}\\
\hline\hline
$\alpha$&ML&a&b&c&$\sigma_{a}$&$\sigma_{b}$&$\sigma_{c}$&$R^2$\\
\hline
F\\
\hline
\endfirsthead
\caption{continued.}\\
\hline\hline
$\alpha$&ML&a&b&c&$\sigma_{a}$&$\sigma_{b}$&$\sigma_{c}$&$R^2$\\
\hline
\endhead
\hline
\endfoot
1.5&A&-3.52&-3.78&3.19&0.04&0.03&0.06&0.998\\
1.5&B&-3.45&-3.76&3.27&0.03&0.03&0.06&0.998\\
1.5&C&-3.27&-3.71&3.18&0.03&0.02&0.05&0.998\\
1.7&A&-3.61&-3.94&3.42&0.09&0.06&0.15&0.999\\
1.7&B&-3.65&-3.91&3.62&0.08&0.06&0.14&0.998\\
1.7&C&-3.21&-3.69&3.09&0.06&0.04&0.11&0.998\\
1.9&A&-3.33&-3.92&3.05&0.12&0.05&0.19&0.999\\
1.9&B&-3.24&-3.93&3.14&0.07&0.01&0.12&0.999\\
1.9&C&-2.89&-3.81&2.81&0.03&0.02&0.06&0.999\\
\hline
FO\\
\hline
1.5&A&-3.53&-3.95&2.48&0.04&0.02&0.06&0.999\\
1.5&B&-3.49&-3.96&2.63&0.06&0.04&0.11&0.999\\
1.5&C&-3.45&-3.97&2.80&0.08&0.07&0.16&0.999\\
1.7&A&-3.49&-3.90&2.38&0.06&0.03&0.11&0.999\\
\hline
\end{longtable}
\end{ThreePartTable}

\begin{ThreePartTable}
\begin{longtable}{ccccccc}
\caption{\label{w_f_fo} PW coefficients ($<W>=<G> -1.9 <G_{BP}-G_{RP}> = a+b\log P$) for F and FO Galactic Cepheids derived by adopting A, B , C ML relations and , $\alpha = 1.5$, $\alpha = 1.7$ and $\alpha = 1.9$ in the \textsl{Gaia} filters.}\\
\hline\hline
$\alpha$&ML&a&b&$\sigma_{a}$&$\sigma_{b}$&$R^2$\\
\hline
F\\
\hline
\endfirsthead
\caption{continued.}\\
\hline\hline
$\alpha$&ML&a&b&$\sigma_{a}$&$\sigma_{b}$&$R^2$\\
\hline
\endhead
\hline
\endfoot
1.5&A&-2.73&-3.26&0.04&0.03&0.995\\
1.5&B&-2.68&-3.18&0.05&0.03&0.994\\
1.5&C&-2.56&-3.16&0.07&0.05&0.992\\
1.7&A&-2.75&-3.37&0.06&0.05&0.998\\
1.7&B&-2.68&-3.20&0.03&0.04&0.996\\
1.7&C&-2.54&-3.23&0.08&0.06&0.997\\
1.9&A&-2.64&-3.52&0.05&0.11&0.999\\
1.9&B&-2.55&-3.44&0.09&0.21&0.999\\
1.9&C&-2.51&-3.11&0.08&0.11&0.999\\
\hline
FO\\
1.5&A&-3.17&-3.80&0.02&0.07&0.999\\
1.5&B&-3.02&-3.85&0.02&0.13&0.996\\
1.7&A&-3.24&-3.96&0.05&0.28&0.999\\
\hline
\end{longtable}
\end{ThreePartTable}

\section{Theoretical versus \textsl{Gaia Data Release 2} parallaxes}
In this section we perform a comparison between the individual
theoretical parallaxes based on the PW relations\footnote{We do not adopt the theoretical PLC relations because they require
a correction for the individual reddening of the observed Cepheid.} in the \textsl{Gaia} filters
and the observed Classical F and FO Cepheids parallaxes taken from
the recent catalog made by \citet{rip2019}. In their work Ripepi and
collaborators reclassify the DR2 Galactic Cepheids and provide
accurate PL and PW relations in the \textsl{Gaia} passbands. To ensure a good
astrometry we chose from the sample the Classical F and FO Cepheids
for which the magnitude in G is brighter than 6 mag and the
renormalized unit weight error values (RUWE) defined by \citet{lind2018} is less than 1.4.

\begin{figure}
\centering
\includegraphics[trim=5 0 30 0, angle = 0, width=0.9\textwidth]{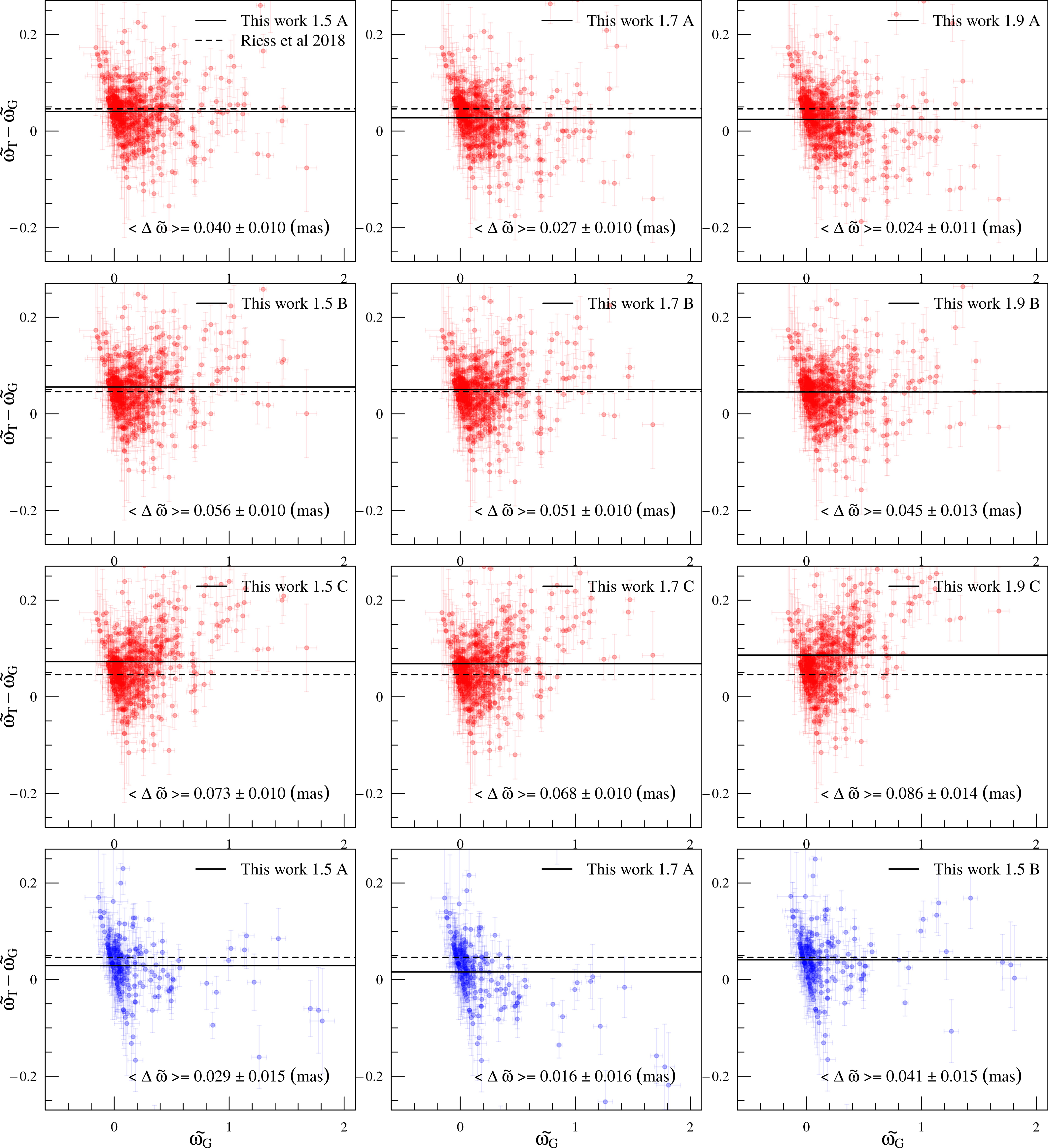}
\caption{Parallax difference $\varpi_{T}$ - $\varpi_{G}$ for F (red points) and FO (blue points) Galactic Cepheids as a function of \textsl{Gaia DR2} parallax $\varpi_{G}$.}
\label{fig:gaia_offset_f_fo}
\end{figure}

The theoretical PW relations derived in the previous subsection are
applied to the observed periods and \textsl{Gaia} magnitudes and colors
reported in the quoted catalog to derive reddening-free individual
distances and in turn theoretical estimates of individual
parallaxes. The latter can
be directly compared with \textsl{Gaia DR2} results as shown in Figure  \ref{fig:gaia_offset_f_fo}.
These plots show the difference between predicted and \textsl{Gaia DR2}
parallaxes versus \textsl{Gaia DR2} parallaxes for the labeled assumptions
concerning the ML relation and the $\alpha$ parameter.
In each panel, the obtained mean offset (solid line) is compared
with the mean offset derived by \citet{riess_zeropoint} (dashed line) and corresponding to $<\Delta\varpi>$=
0.046$\pm$ 0.013 mas, as derived from the HST space astrometric technique. We
notice that this value is reproduced within the errors by our models
apart from a few cases at the brightest luminosity levels (F mode case
C). We also note that variations in the parallax of the order of
$\pm 0.02$ mas at a typical parallax of the order of 0.5 mas implies a
relative parallax error and in turn a relative distance error of
4$\%$. This also reflects on the estimated $H_0$: smaller
 parallaxes by 4$\%$ implies longer distances and in turn smaller values of 
$H_0$ by 4$\%$ that would  be enough to significantly reduce, if not remove the tension. 

\section{Conclusions}
In the context of a theoretical project aimed to investigate the residual systematic effects on the Cepheid-based extragalactic distance scale, a new extended set of nonlinear convective models of Classical Cepheids at solar chemical composition and a wide range of stellar masses and luminosity levels has been computed.
All the predicted pulsation observables for the F and FO models and their dependence on the ML relation and the efficiency of superadiabatic convection have been discussed.
The main results are the following:
\begin{enumerate}
\item As expected, the predicted instability strip gets narrower as the efficiency of superadiabatic convection increases, whereas it does not significantly depend on the assumed ML relation apart from the brighter luminosity levels.
\item Analytical relations connecting the pulsation period of the F and FO models to the intrinsic stellar properties, M, L and $T_{eff}$, have been derived for each assumed mixing length parameter, showing a mild dependence on this value.
\item From the predicted radius curves, mean radii and in turn
theoretical PR and PMR relations have been derived. PR relations have been compared
with similar relations in the literature, showing a good agreement. Moreover, we confirm the results by \citet{bcm1998} for which the PR and PMR relations do not vary considerably with the different assumptions of the ML relation.
\item The obtained bolometric light curves are sensitive to the value of the mixing length parameter with the amplitude decreasing as the efficiency of superadiabatic convection increases, whereas the dependence on the ML relation is much less important.
\item From this set of models the first atlas of theoretical light curves of F and FO Galactic Cepheids converted in the \textsl{Gaia} filters is provided and it shall be made available to the scientific community.
\item The obtained mean magnitudes and colors are used to derive the first theoretical Cepheid PLC and PW relations in the \textsl{Gaia} filters.
\end{enumerate}
Finally the above derived relations have been applied to Galactic Cepheids data in the \textsl{Gaia DR2} database to derive theoretical individual parallaxes which have been compared with the \textsl{Gaia DR2} ones.
In particular, we find that the mean offset derived by \citet{riess_zeropoint} and corresponding to $<\Delta\varpi>$=
0.046$\pm$0.013 mas, is reproduced within the errors by our models
apart from a few cases at the brightest luminosity levels (F mode case
C). To quantify such an offset and its dependence on the physical and
numerical assumptions is crucial to understand and try to reduce the
Hubble constant tension. In particular, a variation in the parallax of the order of
$\pm 0.02$ mas at a typical parallax of the order of 0.5 mas implies a
relative parallax error and in turn a relative error on $H_0$ of
4$\%$. 

\acknowledgments
We thank the Referee for the detailed suggestions and comments. The received feedback significantly improve the quality of the manuscript.
We acknowledge Istituto Nazionale di Fisica Nucleare (INFN), Naples section, specific initiative QGSKY. 
This work has made use of data from the European
Space Agency (ESA) mission \textsl{Gaia} (https://www.cosmos.esa.int/gaia),
processed by the \textsl{Gaia} Data Processing and Analysis Consortium (DPAC, https:
//www.cosmos.esa.int/web/gaia/dpac/consortium). Funding for the
DPAC has been provided by national institutions, in particular the institutions
participating in the Gaia Multilateral Agreement. In particular, the Italian participation
in DPAC has been supported by Istituto Nazionale di Astrofisica (INAF)
and the Agenzia Spaziale Italiana (ASI) through grants I/037/08/0, I/058/10/0, 2014-025-R.0, 2014-025-R.1.2015 and 2018-24-HH.0 to INAF (PI M.G. Lattanzi). We acknowledge partial financial support from 'Progetto Premiale' MIUR MITIC (PI B. Garilli). This work has been partially supported by the INAF Main Stream SSH program, 1.05.01.86.28 and has made use of the
VizieR database, operated at CDS, Strasbourg, France.

\bibliography{giu}{}
\bibliographystyle{aasjournal}

\end{document}